\documentclass[12pt,preprint]{aastex}

\begin{document}
\title{Abundances of Sr, Y, and Zr in Metal-Poor Stars and Implications for 
Chemical Evolution in the Early Galaxy}
\author{Y.-Z. Qian\altaffilmark{1} and G. J. Wasserburg\altaffilmark{2}}
\altaffiltext{1}{School of Physics and Astronomy, University of
Minnesota, Minneapolis, MN 55455; qian@physics.umn.edu.}
\altaffiltext{2}{The Lunatic Asylum, Division of Geological and Planetary
Sciences, California Institute of Technology, Pasadena, CA 91125; 
gjw@gps.caltech.edu.}

\begin{abstract}
Studies of nucleosynthesis in neutrino-driven winds from nascent
neutron stars show that the elements from Sr through Ag with mass
numbers $A\sim 88$--110 are produced by charged-particle reactions
(CPR) during the $\alpha$-process in the winds. Accordingly, we have 
attributed all these elements in stars of low metallicities
(${\rm [Fe/H]}\lesssim -1.5$) to low-mass and normal 
supernovae (SNe) from progenitors of $\sim 8$--$11\,M_\odot$ and 
$\sim 12$--$25\,M_\odot$, respectively, which leave behind neutron 
stars. Using this rule and attributing all Fe production to normal SNe,
we previously developed a phenomenological
two-component model, which predicts that
${\rm [Sr/Fe]}\geq -0.32$ for all metal-poor stars. 
The high-resolution data now available on Sr abundances in 
Galactic halo stars show that there is a great shortfall of Sr relative to 
Fe in many stars with ${\rm [Fe/H]}\lesssim -3$. 
This is in direct conflict with the above prediction. 
The same conflict also exists for two other CPR elements Y and Zr.
The very low abundances of Sr, Y, and Zr observed
in stars with ${\rm [Fe/H]}\lesssim -3$ thus require a stellar
source that cannot be low-mass or normal SNe. We show that this 
observation requires a stellar source leaving behind black holes 
and that hypernovae (HNe) from progenitors of $\sim 25$--$50\,M_\odot$
are the most plausible candidates. Pair-instability SNe from very massive 
stars of $\sim 140$--$260\,M_\odot$ that leave behind no remnants
are not suitable as they are extremely deficient in producing the 
elements of odd atomic numbers such as Na, Al, K, Sc, V, Mn, and Co 
relative to the neighboring elements of even atomic numbers, but this 
extreme odd-even effect is not observed in the elemental abundance 
patterns of metal-poor stars.

If we expand our previous phenomenological two-component 
model to include three components (low-mass and normal SNe and HNe) 
and use for example, the observed abundances of Ba, Sr, and Fe to 
separate the contributions from these components, we find that 
essentially all of the data are very well described by the new model.
This model provides strong constraints on the evolution
of [Sr/Fe] with [Ba/Fe] in terms of the allowed domain for these
abundance ratios. This model also gives an equally good description 
of the data when any 
CPR element besides Sr (e.g., Y or Zr) or any heavy $r$-process
element besides Ba (e.g., La) is used.
As the stars deficient in Sr, Y, and Zr are dominated by contributions 
from HNe, they define the self-consistent yield pattern of 
that hypothecated source. This inferred HN yield pattern for the low-$A$
elements from Na through Zn ($A\sim 23$--70) including Fe is almost 
indistinguishable from what we had previously attributed to normal SNe. 
As HNe are plausible candidates for the first generation of stars and are 
also known to be
ongoing in the present epoch, it is necessary to re-evaluate the extent 
to which normal SNe are substantial contributors to the Fe inventory of the 
Galaxy. We conclude that HNe are important contributors to the abundances 
of the low-$A$ elements over the history of the universe. 
We estimate that they contributed 
$\sim 24\%$ of the bulk solar Fe inventory while normal SNe contributed 
only $\sim 9\%$ (not the usually assumed $\sim 33\%$). This implies a 
greatly reduced role of normal SNe in the chemical evolution of the 
low-$A$ elements. 
\end{abstract}


\section{Introduction}
In this paper we consider that the elements from Sr through Ag in 
metal-poor stars represent the products of nucleosynthesis 
in neutrino-driven winds from forming neutron stars. This 
approach allows us to obtain information on the stellar sources that 
contributed to the chemical enrichment of the interstellar medium 
(ISM) in the Galaxy and the intergalactic medium (IGM) at early 
and recent  times. We previously proposed a phenomenological 
two-component model (\citealt{qw07}; hereafter QW07)
to account for the abundances of heavy elements in metal-poor stars.
That model focused on the elements commonly considered to be
produced by the generic ``$r$-process''.
It specifically attributed all the elements from Sr through Ag 
in metal-poor stars to the charged-particle reactions (CPR) in the 
neutrino-driven winds from nascent neutron stars and used this as a 
diagnostic of the sources for these CPR elements. 
In contrast, the true $r$-process elements (e.g., Ba
and higher atomic numbers) are produced
by extensive rapid neutron capture. It was
assumed in the two-component model
that Fe was only produced by normal supernovae (SNe) 
from progenitors of $\sim 12$--$25\,M_\odot$, 
which leave behind neutron stars, 
and that the heavy $r$-process elements ($r$-elements) with mass
numbers $A>130$ were formed in low-mass SNe from progenitors
of $\sim 8$--$11\,M_\odot$, which also leave behind neutron stars but 
produce no Fe. Thus the CPR elements would be produced by both 
low-mass and normal SNe and the corresponding yields were 
estimated (QW07). It follows that if these SNe were the only sources, 
then the presence of Fe should always be associated with that of the
CPR elements. A more extensive study of the available observational 
data shows that some low-metallicity stars have Fe but essentially no 
Sr. In particular, \citet{fulbright04} found a star with 
${\rm [Fe/H]}=\log{\rm (Fe/H)}-\log{\rm (Fe/H)}_\odot=-2.88$ and 
$\log\epsilon({\rm Sr})=\log({\rm Sr/H})+12<-2.6$ in the dwarf galaxy
Draco. From the two-component model we would have estimated 
$\log\epsilon({\rm Sr})=-0.28$ for this star, which is far above the 
observational upper limit. These results clearly indicate that if 
the CPR elements are always produced during the formation of
neutron stars, then there must be an additional stellar source 
contributing Fe that does not leave behind neutron stars, or else 
the above model for the production of the CPR elements is in error. 

Utilizing a more extensive 
data base than QW07 and especially treating the data on stars very 
deficient in Sr, Y, and Zr relative to Fe at ${\rm [Fe/H]}<-3$,
the present paper will show that a third source in addition to the two 
sources (low-mass and normal SNe) in the model of QW07 is
required to account for
the elemental abundances in metal-poor stars. It will be argued that 
the third source producing Fe but no CPR elements is most likely
associated with hypernovae (HNe) from progenitors of 
$\sim 25$--$50\,M_\odot$, which leave behind black holes instead
of neutron stars.  It is then shown that essentially all of the stellar data
on elemental abundances at ${\rm [Fe/H]}\lesssim -1.5$ can be 
decomposed in terms of three distinct types of sources. This 
decomposition also identifies a yield pattern for the elements from Na
through Zn including Fe that is attributable to HNe. An important 
conclusion is that this HN yield pattern is almost indistinguishable from 
what is attributed to normal SNe. Further, the discovery of extremely 
energetic HNe associated with gamma-ray bursters (e.g., 
\citealt{galama,iwa98}) 
in the present universe requires that contributions from this source must 
be considered both in early epochs and on to the present. This leads to 
a reassessment of the contributions from different sources to the Galactic 
Fe inventory, which shows that ongoing HNe must play an important role 
and that the usual attribution of $\sim 1/3$ of the solar Fe inventory to 
normal SNe is not valid. 

We aim to present a phenomenological three-component (low-mass and 
normal SNe and HNe) model for the chemical evolution of the early Galaxy 
that may provide a quantitative, self-consistent explanation for many of the 
results from stellar observations. We focus on three groups of elements: 
the low-$A$ elements from Na through Zn ($A \sim 23$--70), the CPR
elements from Sr through Ag ($A \sim 88$--110), and the heavy 
$r$-elements ($A > 130$, Ba and higher atomic numbers). 
In \S\ref{sec-2cm} we give a brief outline of the two-component model
of QW07 with low-mass and normal SNe 
represented by the $H$ and $L$ sources,
respectively. In \S\ref{sec-data} we present the data on abundances 
of Sr and Ba as well as Y and La 
for a large sample of metal-poor stars, and show that 
the two-component model fails at ${\rm [Fe/H]}\lesssim -3$ 
and that an additional source producing Fe but no Sr or
heavier elements is required to account for the data at such low 
metallicities. This source is identified with HNe. It is then shown that 
the extended three-component model with HNe, $H$, and 
$L$ sources gives a good representation of nearly all
the data on the CPR elements Sr, Y, and Zr, 
but leads to the conclusion that the HN yield pattern
is indistinguishable from that of the $L$ source for
all the low-$A$ elements. Considering that HNe not only represent 
the first massive stars (Population III stars) but also must
continue into the present epoch, we reinterpret the yields attributed
to the hypothetical $L$ source as the combined contributions
from normal SNe, which we designate as the $L^*$ source,
and HNe. In \S\ref{sec-3cm} we show that the three-component 
model with HNe, $H$, and $L^*$ sources gives a very good 
representation of essentially all the data on the CPR elements
Sr, Y, and Zr and further discuss the 
characteristics of these sources and their roles in the chemical 
evolution of the universe. We give our conclusions in \S\ref{sec-con}.

\section{The Two-Component Model with the $H$ and $L$ Sources}
\label{sec-2cm}
The two-component model\footnote{The original two-component
model was inspired by the meteoritic data on $^{129}$I and $^{182}$Hf
in connection with the $r$-process. See \citet{wbg} for the requirement
of two distinct types of $r$-process sources based on these data and
QW07 for a review on the development of the two-component model.}
of QW07 was based on the observations of 
elemental abundances in metal-poor stars and a
basic understanding of stellar evolution and nucleosynthesis.
It was directed toward identifying the stellar sources for the heavy
$r$-elements. The following
are the key assumptions and inferences of this model:

 (1) The heavy $r$-elements must be produced by an $H$ source that 
contributes essentially none of the low-$A$ elements including Fe. 
The $H$ source is most likely associated
with low-mass SNe from progenitors of $\sim 8$--$11\,M_\odot$
that undergo O-Ne-Mg core collapse.

(2) The low-$A$ elements are produced by an $L$ source
associated with normal SNe from progentiors of $\sim 12$--$25\,M_\odot$
that undergo Fe core collapse. (It was assumed that this source
provided $\sim 1/3$ of the bulk solar Fe inventory.)

(3) The so-called light ``$r$''-elements from Sr through Ag, 
especially Sr, Y, and Zr, in metal-poor stars must have been produced by 
CPRs in the $\alpha$-process \citep{wh92} that 
occurs as material expands away from a nascent neutron star in a 
neutrino-driven wind (e.g., \citealt{dsw86}). Thus, the CPR elements
are not directly related to the
$r$-process (i.e., they are not the true $r$-elements). 
Instead, their production is a natural consequence of
neutron star formation in low-mass and normal SNe associated with the
$H$ and $L$ sources, respectively, as proposed by QW07. 

The above points were incorporated in the two-component model of QW07
to account for the elemental abundances in metal-poor stars. 
For ${\rm [Fe/H]}\lesssim -1.5$, Type Ia SNe (SNe Ia) associated with 
low-mass stars (typically of several $M_\odot$) in binaries had not contributed 
significantly to the Fe group elements in the ISM.
Similarly, there were no significant contributions to Sr and heavier elements
in the ISM of this early regime from the $s$-process in 
asymptotic giant branch (AGB) stars.
Thus, it was considered in QW07 that for ${\rm [Fe/H]}\lesssim -1.5$, 
the $H$ source is solely responsible for the heavy $r$-elements such as Eu 
and the $L$ source is solely responsible for the low-$A$ elements such as 
Fe while both sources produce the CPR elements. The yield pattern for
the prototypical $H$ source was taken from the data on a star 
(CS~22892--052, \citealt{sneden03}) with extremely high enrichment in 
the heavy $r$-elements relative to the low-$A$ elements. In contrast,
the yield pattern for the prototypical $L$ source was taken from the data 
on a star (HD~122563, \citealt{honda06}) with very little enrichment in the 
heavy $r$-elements relative to the low-$A$ elements and the abundances 
of the latter elements in this star
were attributed to the $L$ source only. For this two-component model,
the (number) abundance of an element E in the ISM at 
${\rm [Fe/H]}\lesssim -1.5$ can be calculated as
\begin{equation}
\left(\frac{\rm E}{\rm H}\right)=\left(\frac{\rm E}{\rm Eu}\right)_H
\left(\frac{\rm Eu}{\rm H}\right)+\left(\frac{\rm E}{\rm Fe}\right)_L
\left(\frac{\rm Fe}{\rm H}\right),
\label{eq-eh}
\end{equation}
where (E/Eu)$_H$ and (E/Fe)$_L$ are the (number)
yield ratios of E to Eu and
Fe for the $H$ and $L$ sources, respectively. Given these yield
ratios, the abundances of all the other elements (relative to hydrogen)
in a star can be 
obtained from the above equation using only the observed abundances 
of Eu and Fe in that star. The results from the above 
model were in good agreement with the data on a large sample of 
metal-poor stars. We note that so long as there are no significant 
$s$-process contributions
to the ISM (which is the case for ${\rm [Fe/H]}\lesssim -1.5$) or the star
has not undergone mass transfer from an AGB companion in a binary,
the element Ba can also be used as a measure of the $r$-process 
contributions. The yield ratios (E/Eu)$_H$ and (E/Fe)$_L$ for the heavy 
$r$-elements
and the CPR elements, as well as the yield ratios (E/Ba)$_H$ for the
CPR elements, are given in Tables~\ref{tab-rhl} and \ref{tab-yhl}.

We emphasize the phenomenological nature of the two-component
model and its extension presented below. As discussed above, the
yield patterns for the $H$ and $L$ sources were taken from the
observed abundance patterns in two template stars. The validity of
the model should be judged by its predictions for the abundances in
other metal-poor stars. Insofar as the predictions agree with the data,
the model can be considered to have identified some key 
characteristics of nucleosynthesis in the relevant stellar sources.
This approach cannot replace the ab initio models of stellar
nucleosynthesis, but is complimentary to the latter.

We note that the production of the low-$A$ elements
including Fe in normal SNe from progentiors of 
$\sim 12$--$25\,M_\odot$ ($L$ source) is demonstrated by extensive 
modeling of SN nucleosynthesis (e.g., \citealt{ww95,tnh96,cl04}), and
so is the production of the CPR elements in the neutrino-driven wind
associated with neutron star formation ($H$ and $L$ sources; e.g., \citealt{meyer92,taka94,woosley94,hoffman97}). 
However, the theoretical yields of the low-$A$ elements, especially
the Fe group,  are subject to 
the many uncertainties in modeling the evolution and explosion of 
massive stars. In the absence of a solid understanding of the SN
mechanism, the explosion is artificially induced and the associated
nucleosynthesis is parametrized by a ``mass cut'' (e.g., \citealt{ww95})
or constrained to fit the yields of $^{56}$Ni inferred from SN light
curves (e.g., \citealt{cl04}). For the CPR elements,
no reliable quantitative yields are yet available. QW07 concluded
that the neutrino-driven wind does not play a significant role in the
production of the heavy $r$-elements and suggested that another
environment with rapid expansion timescales inside low-mass 
($\sim 8$--$11\,M_\odot$) SNe from O-Ne-Mg core collapse ($H$ source) 
is responsible for making these
elements. Subsequent work by \citet{ning07} showed that the propagation
of a fast shock through the surface layers of an O-Ne-Mg core can
provide the conditions leading to the production of the heavy $r$-elements.
However, the required shock speed is not obtained in the current
SN models \citep{janka07} based on the pre-SN structure of a 
$1.38\ M_\odot$ core calculated by 
\citet{nomoto84,nomoto87}. Clearly, more studies of the pre-SN
evolution of O-Ne-Mg cores and their collapse are needed to test
whether the heavy $r$-elements can indeed be produced in the shocked 
surface layers of such cores. In the following we assume that
low-mass SNe from O-Ne-Mg core collapse are the $H$ source solely 
responsible for producing the heavy $r$-elements and that the CPR
elements are produced by both the $H$ and $L$ sources.
At the present time, stellar models cannot calculate the absolute yields 
from first principles for any of the sources discussed above and
some ad hoc parametric treatment is required for modeling the
nucleosynthesis of these sources.

\section{Failure of the Two-Component Model and Requirement of HNe}
\label{sec-data}
\begin{figure}
\includegraphics[angle=270,scale=.30]{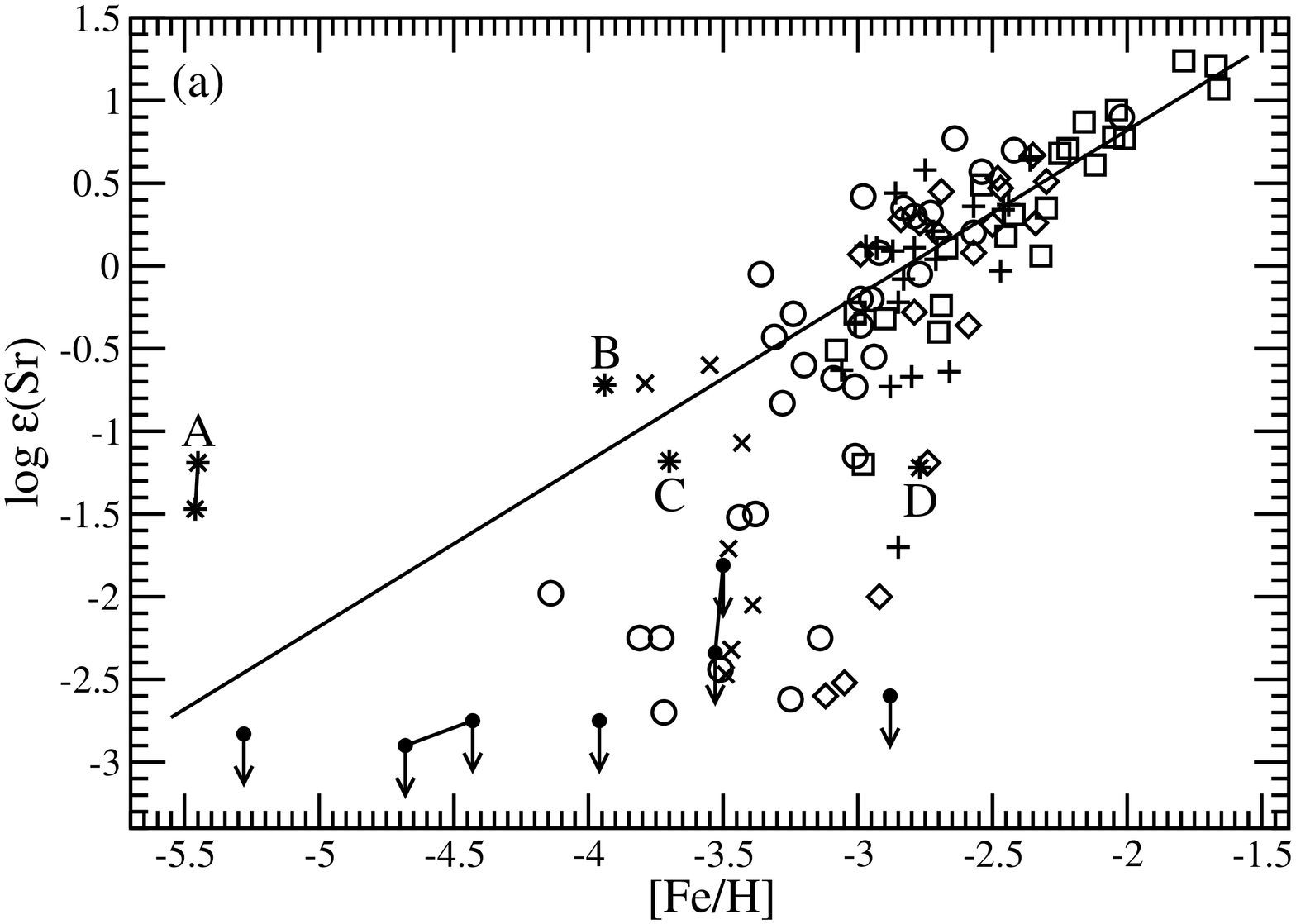}
\includegraphics[angle=270,scale=.30]{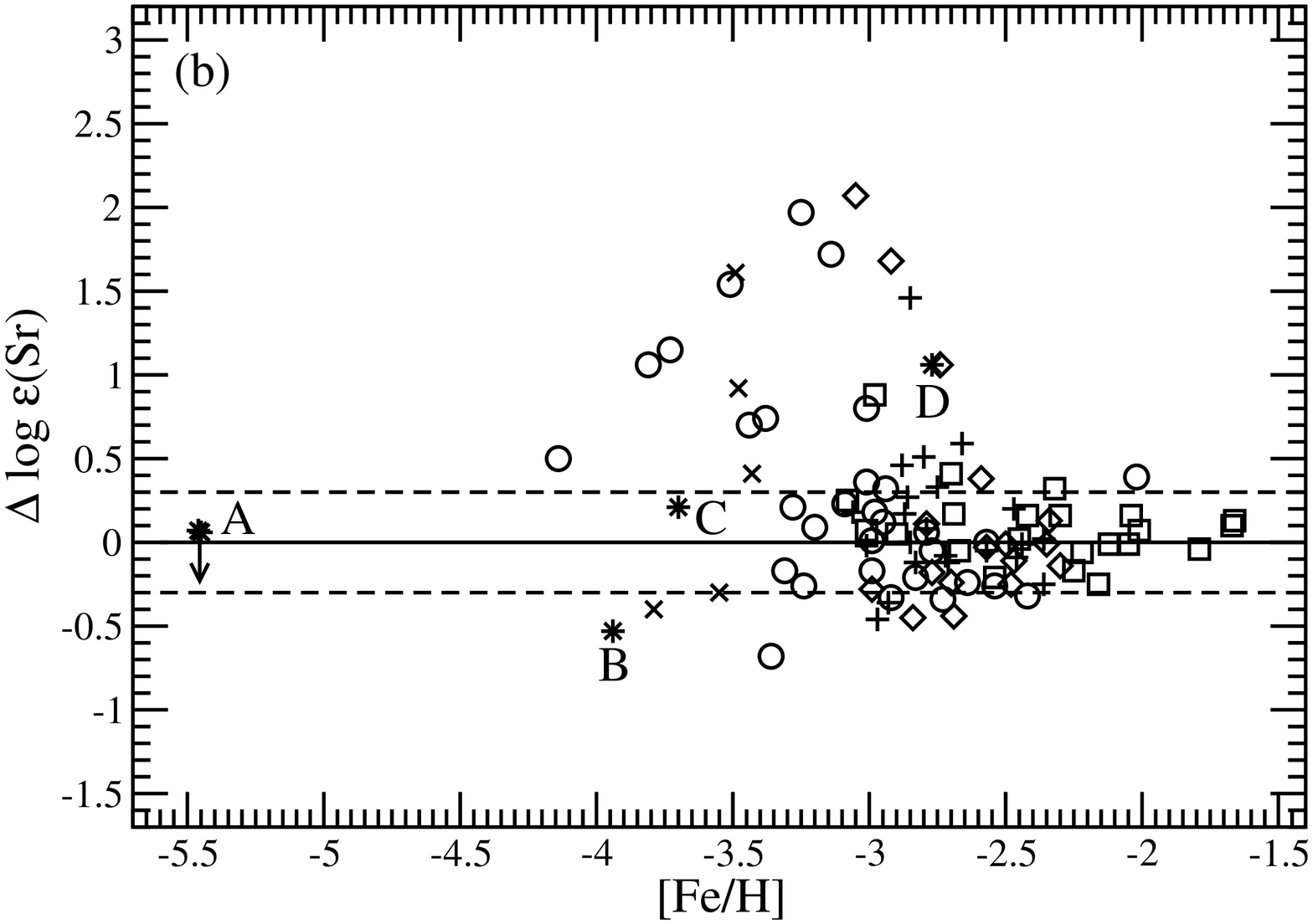}
\caption{(a) High-resolution data on $\log\epsilon({\rm Sr})$ vs. [Fe/H] 
(squares: \citealt{johnson02},
pluses: \citealt{honda04}, diamonds: \citealt{aoki05}, circles:
\citealt{francois07}, crosses: \citealt{cohen08}, asterisks representing
stars with very high C and O abundances and anomalous abundance
patterns of the low-$A$ elements:
\citealt{aoki06} (A, HE~1327--2326);  
\citealt{depagne02} (B, CS~22949--037);
\citealt{aoki02} (C, CS~29498--043);
\citealt{aoki07} (D, BS~16934--002), downward arrows indicating 
upper limits: \citealt{christlieb04,fulbright04,frebel07,cohen07,norris07}).
Symbols connected with a line indicate results for the same star assuming
two different atmospheric models (subgiant vs. dwarf). Typical observational
errors in $\log\epsilon({\rm Sr})$ are $\sim 0.2$--0.3 dex.
The solid line is for an ISM with well-mixed $H$ and $L$ contributions.
The data mostly cluster around this line but drastically depart to low 
$\log\epsilon({\rm Sr})$ values for ${\rm [Fe/H]}\lesssim -3$. 
(b) Comparison of the two-component model of QW07 and the observations
in terms of $\Delta\log\epsilon({\rm Sr})\equiv
\log\epsilon_{\rm cal}({\rm Sr})-\log\epsilon_{\rm obs}({\rm Sr})$ as a function 
of [Fe/H] for those stars shown in (a) that have observed Ba abundances.
In general, the model grossly overestimates the Sr abundance below 
${\rm [Fe/H]}\sim -2.7$. However, the calculated Sr abundance for 
HE~1327--2326 with ${\rm [Fe/H]}=-5.45$ (asterisk A) using the
upper limit on its Ba abundance appears to be in good agreement with
its observed Sr abundance. Measurement of the exact Ba abundance
in this star will provide an extremely important test of the model.}
\label{fig-esr}
\end{figure}

In Figure~\ref{fig-esr}a we show the data on 
$\log\epsilon({\rm Sr})$ from an extensive set of 
the available high-resolution observations over the wide range 
of $-5.5\lesssim {\rm [Fe/H]}\lesssim -1.5$ (squares: \citealt{johnson02};
pluses: \citealt{honda04}; diamonds: \citealt{aoki05}; circles:
\citealt{francois07}; crosses: \citealt{cohen08}; asterisks:
\citealt{depagne02,aoki02,aoki06,aoki07}; downward arrows indicating
upper limits: \citealt{christlieb04,fulbright04,frebel07,cohen07,norris07}).
All the data are for stars in the Galactic halo except for the downward 
arrow at ${\rm [Fe/H]}=-2.88$, which is for a star 
(Draco 119, \citealt{fulbright04}) in the dwarf galaxy Draco.

We now use the two-component model of QW07 to analyze the data 
shown in Figure~\ref{fig-esr}a. We first apply equation~(\ref{eq-eh}) 
to calculate the $H$ and $L$ contributions to the solar Sr abundance 
assuming that the $H$ source provided all of the solar Eu abundance 
and the $L$ source provided 1/3 of the solar Fe abundance. 
Further assuming that the sun 
represents the sampling of a well-mixed ISM, we can show that such 
an ISM has [Sr/Fe]$_{\rm mix}=-0.10$ resulting from the mixing of 
$H$ and $L$ contributions only (see Appendix~\ref{sec-app1} and
Table~\ref{tab-mix}). This Sr/Fe ratio corresponds to
\begin{equation}
\log\epsilon({\rm Sr})={\rm [Fe/H]}+2.82,
\end{equation}
shown as the solid line in Figure~\ref{fig-esr}a. It can be
seen from this figure that the bulk of the data lie close to the solid line, 
but for ${\rm [Fe/H]}\lesssim -3$, almost all of the data depart greatly 
from this line. 

There is a lack of Eu data for 
many stars with $[{\rm Fe/H}]<-3$. As Ba data are more readily 
available for such stars, we use Ba instead of Eu
as the index heavy $r$-element
to identify the contributions from the $H$ source
(this is robust so long as there are no $s$-process contributions). 
Then equation~(\ref{eq-eh}) can be rewritten for Sr as
\begin{equation}
\left(\frac{\rm Sr}{\rm H}\right)=\left(\frac{\rm Sr}{\rm Ba}\right)_H
\left(\frac{\rm Ba}{\rm H}\right)+\left(\frac{\rm Sr}{\rm Fe}\right)_L
\left(\frac{\rm Fe}{\rm H}\right).
\label{eq-srh}
\end{equation}
The yield ratios (E/Ba)$_H$ and (E/Fe)$_L$ for Sr and other 
CPR elements are given in Table~\ref{tab-yhl}. Using
these yield ratios and the above equation, we calculate the 
$\log\epsilon_{\rm cal}{\rm (Sr)}$ 
values for those stars shown in Figure~\ref{fig-esr}a that have 
observed Ba and Fe abundances. The differences 
$\Delta\log\epsilon({\rm Sr})\equiv\log\epsilon_{\rm cal}{\rm (Sr)}-
\log\epsilon_{\rm obs}{\rm (Sr)}$ between the calculated and
observed values are shown in Figure~\ref{fig-esr}b. 
Note that for ${\rm [Fe/H]}> -2.7$, the agreement between
the model predictions and the data is very good. However, for 
${\rm [Fe/H]}\lesssim -2.7$, there is great discrepancy in the
sense that the calculated $\log\epsilon_{\rm cal}{\rm (Sr)}$ 
values for many stars far exceed the observed values.
It is this discrepancy that we will focus on in this paper.

The large disgreement between the model predictions and the 
data for ${\rm [Fe/H]}\lesssim -2.7$ shown in Figure~\ref{fig-esr}b
is caused by assigning all the Fe to the $L$ source.
If there is an additional source producing Fe but no Sr or 
heavier elements at such low metallicities, then 
equation~(\ref{eq-srh}) overestimates the Sr abundances.
The requirement of such a source
can also be seen from the Sr/Fe ratios for the stars. The yield
ratio (Sr/Fe)$_L$ corresponds to [Sr/Fe]$_L=-0.32$ (see 
Table~\ref{tab-mix}). 
Of the $H$ and $L$ sources, both produce Sr but only the latter 
can produce Fe. Thus any mixture of 
the contributions from these two sources should have 
${\rm [Sr/Fe]}\geq -0.32$. Figure~\ref{fig-srfe} shows [Sr/Fe] 
vs. [Ba/Fe] for those stars in Figure~\ref{fig-esr} that have
observed Ba abundances or upper limits. It can be seen that
many stars have ${\rm [Sr/Fe]}\ll -0.32$ and quite a few have 
${\rm [Sr/Fe]}\lesssim -2$. These observations are in direct conflict 
with the two-component model and can only be accounted for if 
there is an additional source for Fe (and the associated 
elements) that produce none or very little of Sr and heavier 
elements. If we expand the framework of QW07 to include 
this third source, then a self-consistent interpretation of all the data 
may be possible.                                                    

\subsection{Effects of the Third Source}
\label{sec-p}
In the extended model including the third source in addition to
the $H$ and $L$ sources,
only a fraction $f_{{\rm Fe},L}$ of the Fe is produced by the $L$ source.
Then equation~(\ref{eq-srh}) becomes
\begin{equation}
\left(\frac{\rm Sr}{\rm H}\right)=\left(\frac{\rm Sr}{\rm Ba}\right)_H
\left(\frac{\rm Ba}{\rm H}\right)+\left(\frac{\rm Sr}{\rm Fe}\right)_L
\left(\frac{\rm Fe}{\rm H}\right)f_{{\rm Fe},L}.
\label{eq-srhf}
\end{equation}
For $f_{{\rm Fe},L}=1$ the extended model reduces to the two-component
model. For $f_{{\rm Fe},L}=0$ only the $H$ source and the third source are 
relevant with the latter being the sole contributor of the low-$A$ elements 
such as Fe and the former being the sole contributor of Sr and heavier 
elements. Equation~(\ref{eq-srhf}) can be rewritten as
\begin{equation}
{\rm [Sr/Fe]}=\log\left(10^{{\rm [Sr/Ba]}_H+{\rm [Ba/Fe]}}+
f_{{\rm Fe},L}\times 10^{{\rm [Sr/Fe]}_L}\right).
\label{eq-srfe}
\end{equation}
Note that the above equation represents very strong constraints on 
the evolution of [Sr/Fe] with [Ba/Fe] in the extended model. Whether
a system starts with the initial composition of big bang debris or with
an initial state inside the region defined by the curves representing
equation~(\ref{eq-srfe}) for 
$f_{{\rm Fe},L}=0$ and 1, it cannot have [Sr/Fe] and 
[Ba/Fe] values outside this region upon further evolution so long as
only the $H$ and $L$ sources and the third source contribute metals.

\begin{figure}
\includegraphics[angle=270,scale=.50]{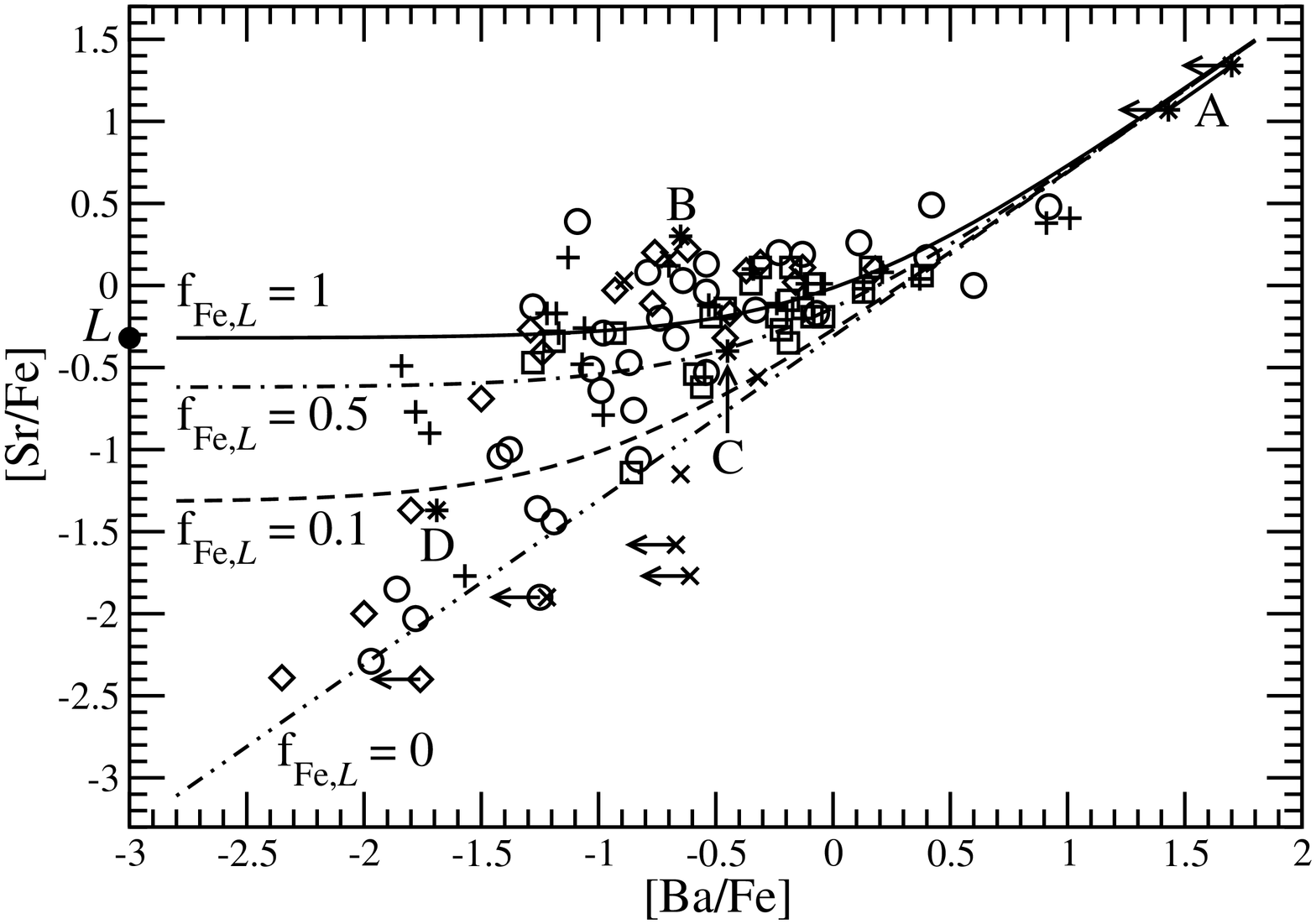}
\caption{Evolution of [Sr/Fe] with [Ba/Fe]. Data symbols are the same as
in Figure~\ref{fig-esr} except that the left-pointing arrows indicate the upper 
limits on [Ba/Fe]. Typical observational errors in [Sr/Fe] and [Ba/Fe] are
$\sim 0.1$--0.25 dex. The curves show 
${\rm [Sr/Fe]}=\log\left(10^{{\rm [Sr/Ba]}_H+{\rm [Ba/Fe]}}+ f_{{\rm Fe},L}
\times 10^{{\rm [Sr/Fe]}_L}\right)$ based on the three-component model
with the $H$ and $L$ sources and a third source (HNe)
for $f_{{\rm Fe},L}=0$ (dot-dot-dashed), 0.1 (dashed), 0.5 (dot-dashed),
and 1 (solid). The parameter $f_{{\rm Fe},L}$ is the fraction of Fe 
contributed by the $L$ source ($f_{{\rm Fe},L}=0$ corresponds to
all the Fe being from the third source). The filled circle labeled ``$L$''
indicates the value of [Sr/Fe]$_L=-0.32$ for the $L$ source. 
Almost all of the data lie within 
the allowed region of the model. Note the 
presence of quite a few data on the curve for $f_{{\rm Fe},L}=0$
as well as the abundant data near the curve for $f_{{\rm Fe},L}=1$.}
\label{fig-srfe}
\end{figure}

The curves representing equation~(\ref{eq-srfe}) for 
$f_{{\rm Fe},L}=0$, 0.1,
0.5, and 1 are shown along with the data on [Sr/Fe] vs. [Ba/Fe] in
Figure~\ref{fig-srfe}. Most of the data lie between the curves for
$f_{{\rm Fe},L}=0$ and 1 with a clustering of the data around the 
curve for $f_{{\rm Fe},L}=1$. There are also quite a few data on the
curve for $f_{{\rm Fe},L}=0$. Some data lie distinctly above the curve 
for $f_{{\rm Fe},L}=1$. This could be partly due to observational 
uncertainties. We also note that all SNe associated with the $H$ 
and $L$ sources are assumed here to have fixed yield patterns.
If there were variations in the yield ratios by
a factor of several, then some ``forbidden'' region above the curve
for $f_{{\rm Fe},L}=1$ would be accessible. In \S\ref{sec-3cm} we
will show that with a reinterpretation of the $L$ source there is no
longer a need to call upon such variabilities. In any case,
we consider that the comparison between the theoretical model 
curves for $f_{{\rm Fe},L}=0$ to 1 and the data shown in 
Figure~\ref{fig-srfe} justifies the extended model where 
a third source is producing Fe but no Sr or heavier elements.

\begin{figure}
\includegraphics[angle=270,scale=.50]{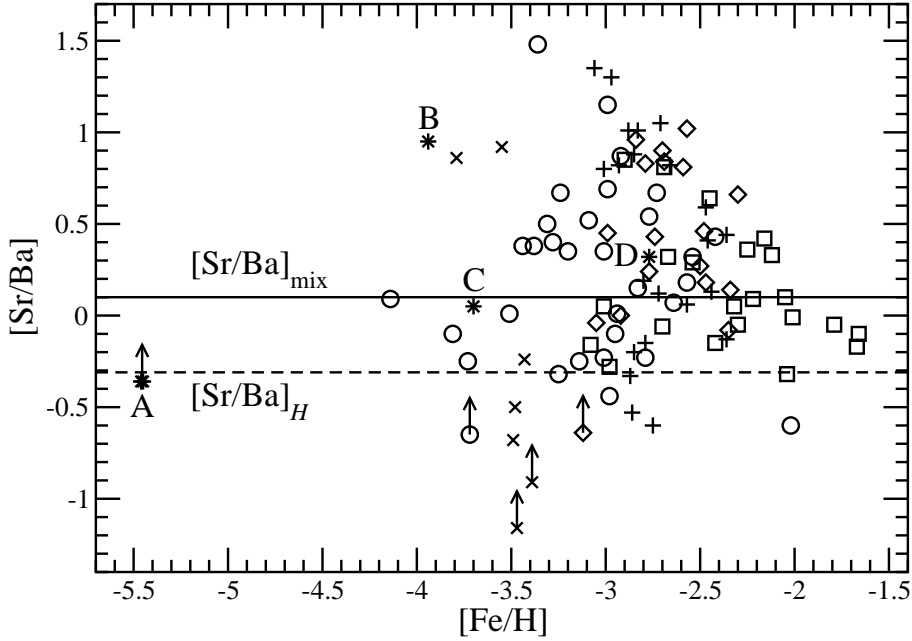}
\caption{Data on [Sr/Ba] vs. [Fe/H]. Symbols are the same as in
Figure~\ref{fig-esr} except that the upward arrows represent lower
limits on [Sr/Ba]. Typical observational errors in [Sr/Ba] are
$\sim 0.2$--0.3 dex. The dashed line shows the lower bound of 
[Sr/Ba]$_H=-0.31$ for pure $H$ contributions and the solid line
shows the value of [Sr/Ba]$_{\rm mix}=0.10$ for an ISM with well-mixed
$H$ and $L$ contributions. Data above the solid line represent higher 
proportions of $L$ contributions than in the well-mixed case. Note that
considering observational uncertainties, there are no serious exceptions 
to the rules of Fe, Sr, and Ba production for the $H$ and $L$ 
sources and the third source (HNe) in the three-component model.}
\label{fig-srba}
\end{figure}

With no production of Sr or heavier elements assigned to the third 
source, the Sr/Ba ratio is determined exclusively by the $H$ and $L$ 
sources. As both these sources produce Sr but only the $H$ source
can produce Ba, any mixture of the contributions from these two sources 
should have ${\rm [Sr/Ba]}\geq {\rm [Sr/Ba]}_H$. Figure~\ref{fig-srba}
shows the data on [Sr/Ba] vs. [Fe/H] along with two reference lines, 
one corresponding to [Sr/Ba]$_H=-0.31$ (see Table~\ref{tab-mix})
and the other to [Sr/Ba]$_{\rm mix}=0.10$ for an ISM with well-mixed 
$H$ and $L$ contributions (see Appendix~\ref{sec-app1} and 
Table~\ref{tab-mix}). 
An excess of $L$ over $H$ contributions relative to the well-mixed
case displaces [Sr/Ba] above the line for [Sr/Ba]$_{\rm mix}$. It can be 
seen from Figure~\ref{fig-srba} that almost all of the data are compatible 
with the lower bound of ${\rm [Sr/Ba]}\geq {\rm [Sr/Ba]}_H$ and that a 
substantial fraction of the stars did not sample a well-mixed ISM. 
Excluding the lower limits, we note that some data lie below the line 
for [Sr/Ba]$_H$. However, the deviation below [Sr/Ba]$_H$ is
$\lesssim 0.4$~dex, which is comparable to the observational 
uncertainties\footnote{Four of the data points are repeated
measurements of well-studied stars with large $r$-process 
enrichments: the plus at ${\rm [Fe/H]}=-2.86$, ${\rm [Sr/Ba]}=-0.53$
and the circle at ${\rm [Fe/H]}=-2.98$, ${\rm [Sr/Ba]}=-0.44$
for CS~22892--052, the plus at ${\rm [Fe/H]}=-2.75$, 
${\rm [Sr/Ba]}=-0.60$ for CS~31082--001, and the circle at 
${\rm [Fe/H]}=-2.02$, ${\rm [Sr/Ba]}=-0.60$ for BD~$+17^\circ 3248$.
Observational studies focused on these stars give
${\rm [Sr/Ba]}=-0.31$ (CS~22892--052, \citealt{sneden03}), 
$-0.43$ (CS~31082--001, \citealt{hill02}), and $-0.16$
(BD~$+17^\circ 3248$, \citealt{cowan02}). We note that having some 
$s$-process contributions to Ba would also lower [Sr/Ba].}
and does not represent serious violation of the lower bound.
It is important to note that the four stars shown as asterisks 
A, B, C, and D in Figures~\ref{fig-esr}, \ref{fig-srfe}, and \ref{fig-srba}
appear to be well behaved in terms of Sr and Ba although
they have very high abundances of C and O and anomalous 
abundance patterns of the low-$A$ elements (see \S\ref{sec-pan}).

\begin{figure}
\includegraphics[angle=270,scale=.30]{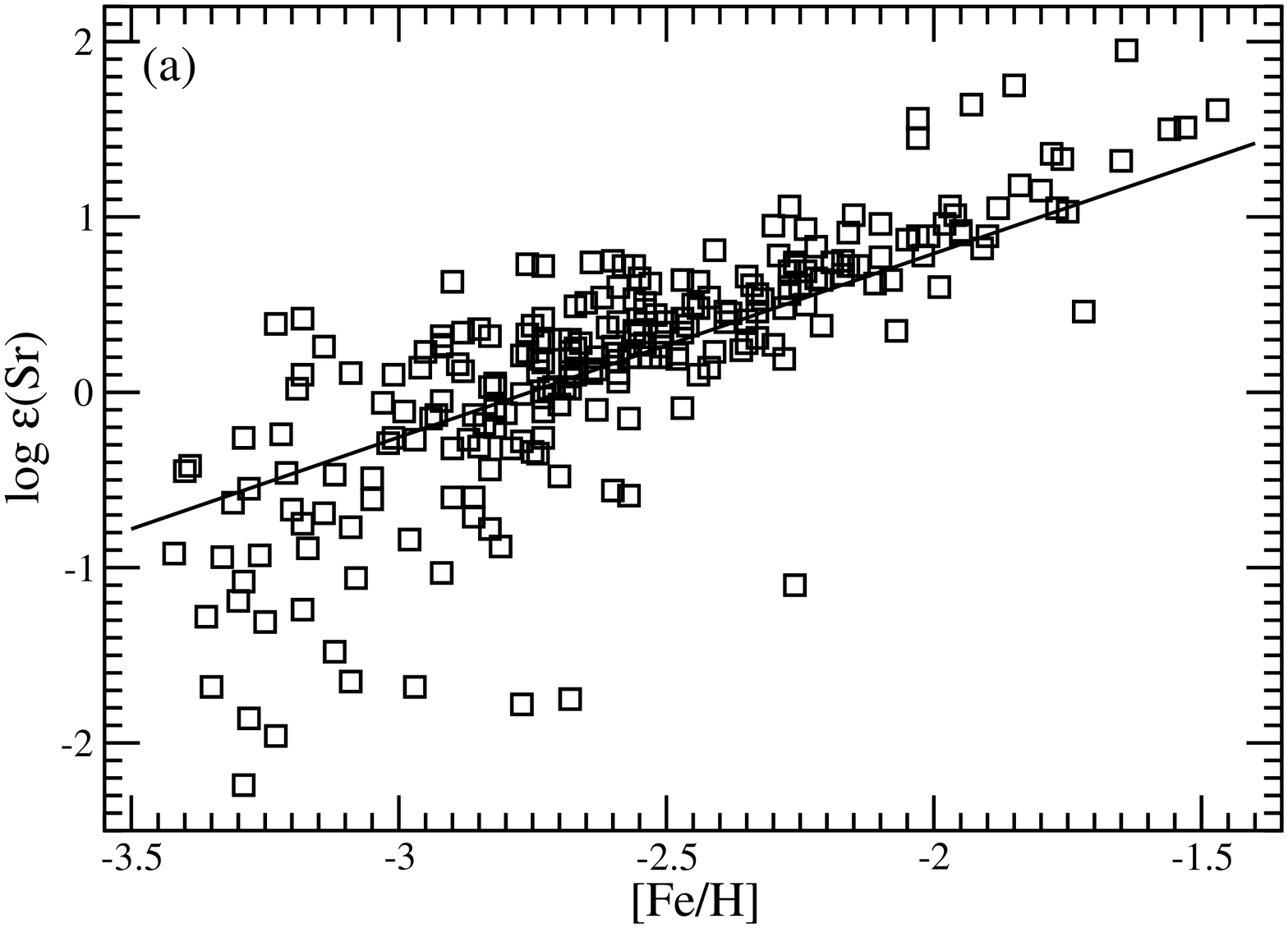}
\includegraphics[angle=270,scale=.30]{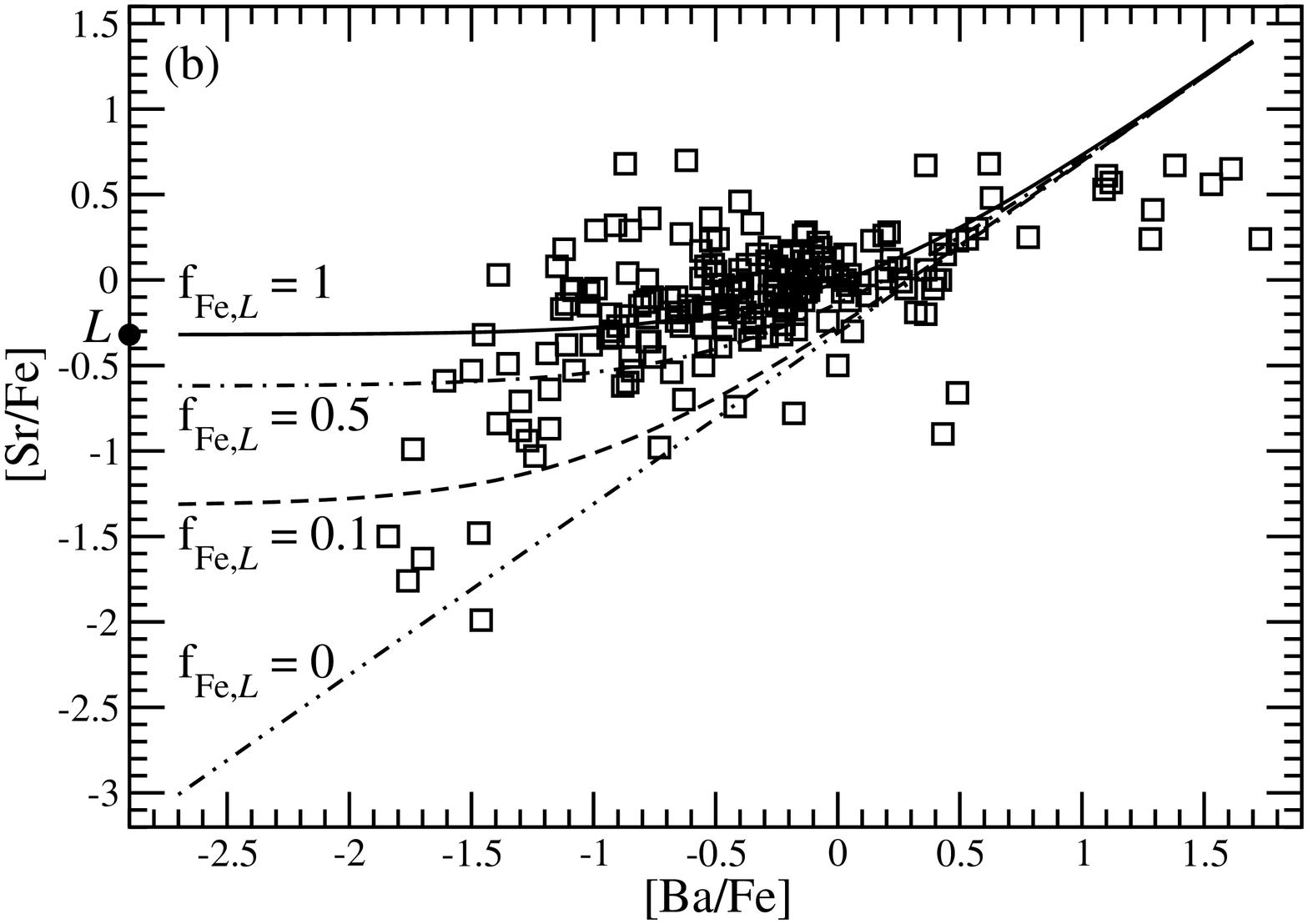}
\caption{(a) Medium-resolution data on $\log\epsilon({\rm Sr})$ vs. [Fe/H]
from the HERES survey \citep{barklem05}. Typical observational errors in 
$\log\epsilon({\rm Sr})$ are $\sim 0.3$ dex. The data in general follow the 
same distribution as presented in Figure~\ref{fig-esr}a for the 
high-resolution data, where the same solid line for an ISM with
well-mixed $H$ and $L$ contributions is also shown. The majority of the 
data cluster around the solid line but there is a great dispersion 
below ${\rm [Fe/H]}\sim -2.5$. (b) Evolution of [Sr/Fe] with [Ba/Fe] for the 
HERES sample. Typical observational errors in [Sr/Fe] and [Ba/Fe] are
$\sim 0.3$ dex. The data distribution is again quite similar to the case 
for the high-resolution data presented in Figure~\ref{fig-srfe}, where
the same curves are shown. A number of
data lie far to the right of and below the curve for $f_{{\rm Fe},L}=0$.
We consider that the corresponding stars (HE~0231--4016,
HE~0305--4520, HE~0430--4404, HE~1430--1123, HE~2150--0825,
HE~2156--3130, HE~2227--4044, and HE~2240--0412)
may have received large
$s$-process contributions. This can be tested by high-resolution 
observations covering more elements heavier than Ba.}
\label{fig-heres}
\end{figure}

To further test the robustness of the extended model including 
the third source,
we have carried out a similar analysis of the medium-resolution data
from the HERES survey of metal-poor stars \citep{barklem05}.
This sample contains 253 stars of which eight have neither Sr nor Ba 
data. Five of the remaining stars were clearly recognized from their
Ba/Eu ratios as having dominant $s$-process contributions 
\citep{barklem05} and are excluded. This leaves 240 stars to be analyzed 
here. The data on $\log\epsilon({\rm Sr})$ vs. [Fe/H] are shown in 
Figure~\ref{fig-heres}a analogous to Figure~\ref{fig-esr}a. It can be seen 
that the bulk of the data again cluster around the line for an ISM with 
well-mixed $H$ and $L$ contributions but there are again many stars 
with ${\rm [Fe/H]}\lesssim -3$ showing a great deficiency in Sr. 
The description for the evolution of [Sr/Fe] with [Ba/Fe] 
by the extended model is compared with the medium-resolution
data in Figure~\ref{fig-heres}b.
In general, these medium-resolution data are in accord with the results 
presented above for the high-resoltuion data shown in Figure~\ref{fig-srfe}, 
but with some exceptions. 
There are a number of data that lie well above the upper bound for
$f_{{\rm Fe},L}=1$ (e.g., the data at ${\rm [Ba/Fe]}=-0.87$, 
${\rm [Sr/Fe]}=0.68$ and ${\rm [Ba/Fe]}=-0.62$, ${\rm [Sr/Fe]}=0.70$
representing HE~0017--4838 and HE~1252--0044, respectively). This
may be partly due to observational uncertainties, but it will be shown
in \S\ref{sec-3cm} that a reinterpretation of the $L$ source raises the
upper bound above essentially all the data. There are also a number
of data that lie far to the right of and below the lower bound for 
$f_{{\rm Fe},L}=0$. We consider that the corresponding stars 
most plausibly have large $s$-process contributions to the Ba  
(these stars are: HE~0231--4016,
HE~0305--4520, HE~0430--4404, HE~1430--1123, HE~2150--0825,
HE~2156--3130, HE~2227--4044, and HE~2240--0412).
This explanation can be tested by high-resolution observations covering 
more elements heavier than Ba. While not showing 
as clear-cut a case as the high-resolution data, the bulk of the HERES
data appear to be in broad accord with the requirement of a third 
source producing Fe but no Sr or heavier elements as presented 
above. 

\subsection{Requirement of a Third Source from Data on Y and La 
as well as Zr and Ba}
\label{sec-yla}

\begin{figure}
\includegraphics[angle=270,scale=.30]{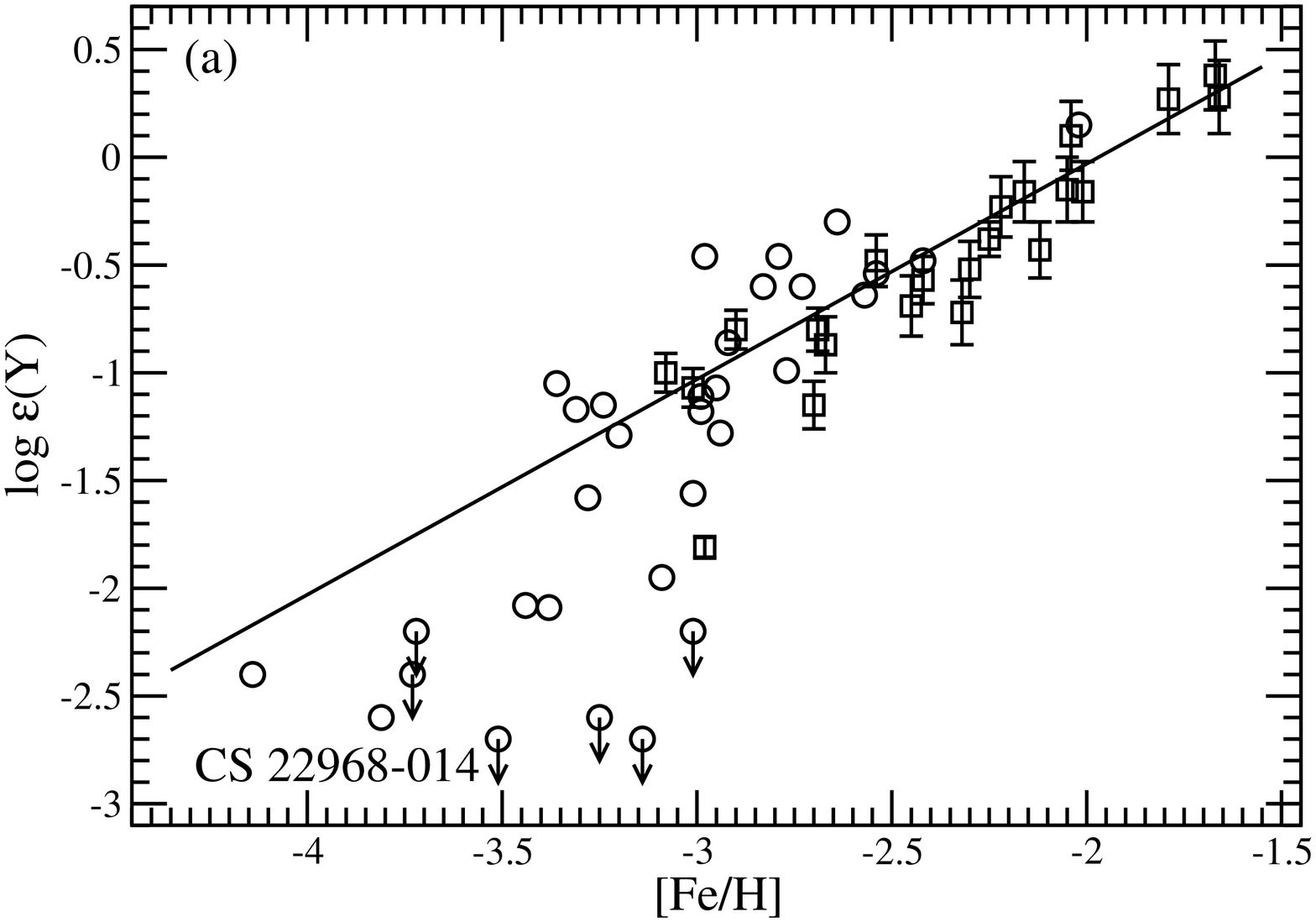}
\includegraphics[angle=270,scale=.30]{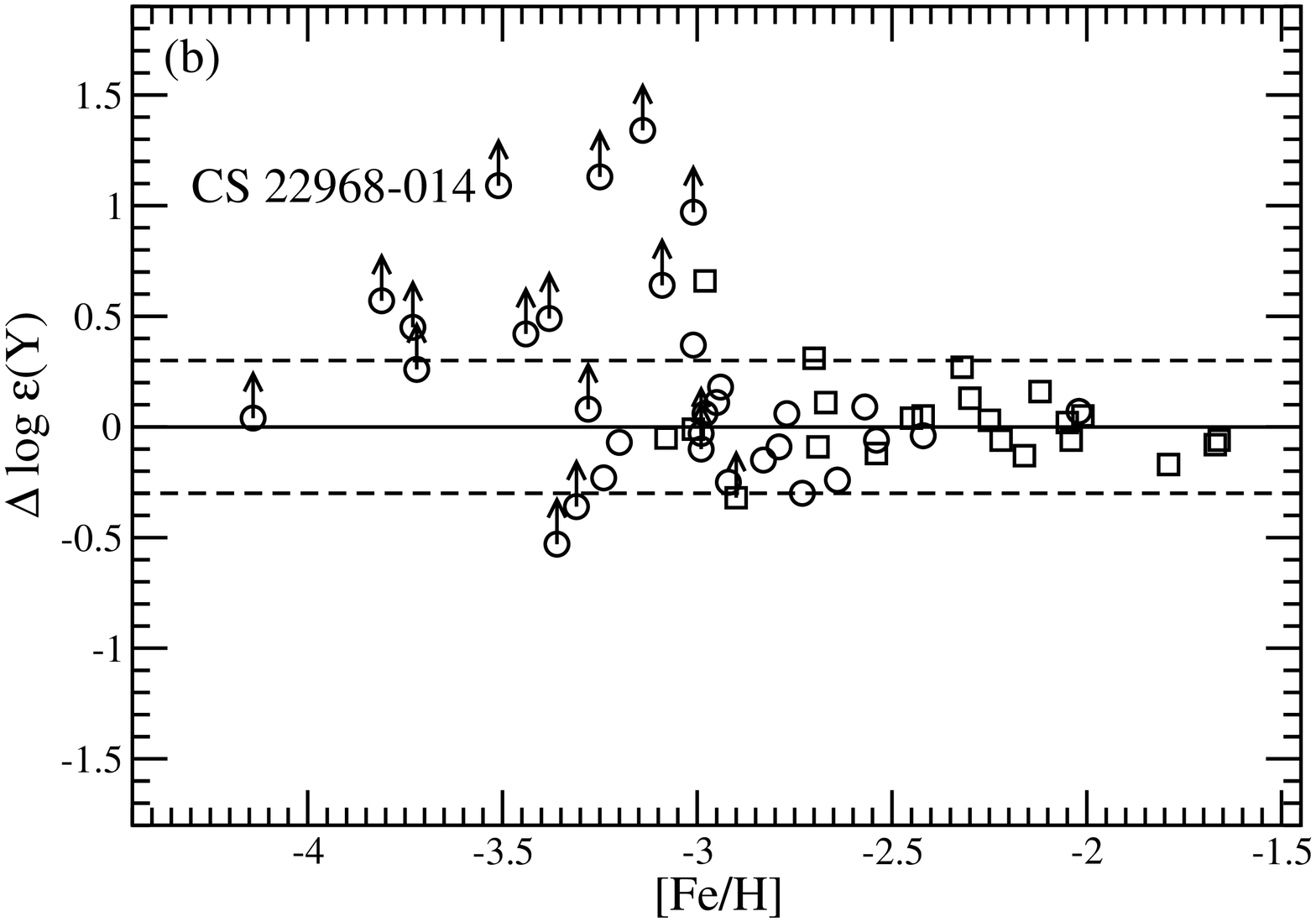}
\includegraphics[angle=270,scale=.30]{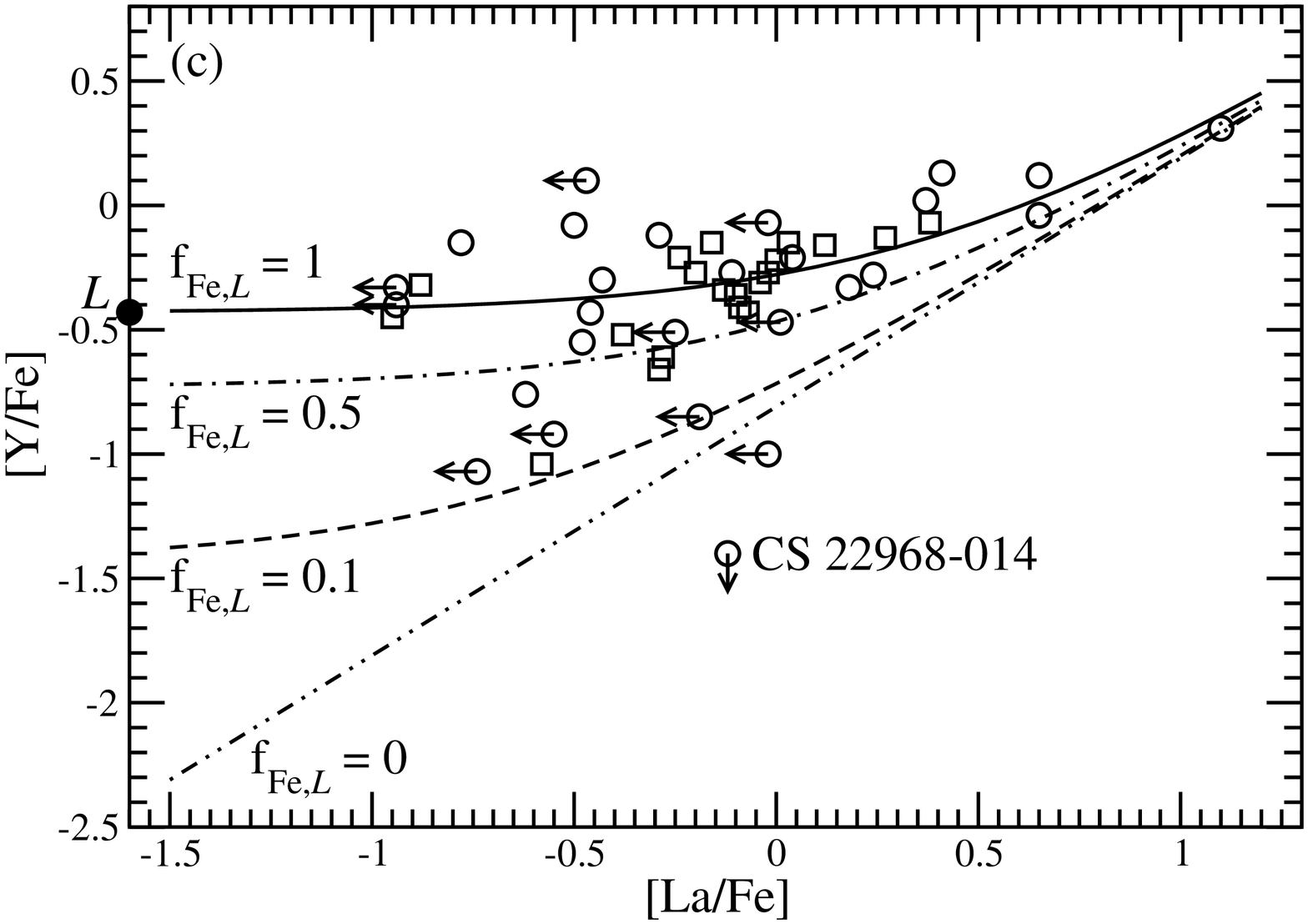}
\includegraphics[angle=270,scale=.30]{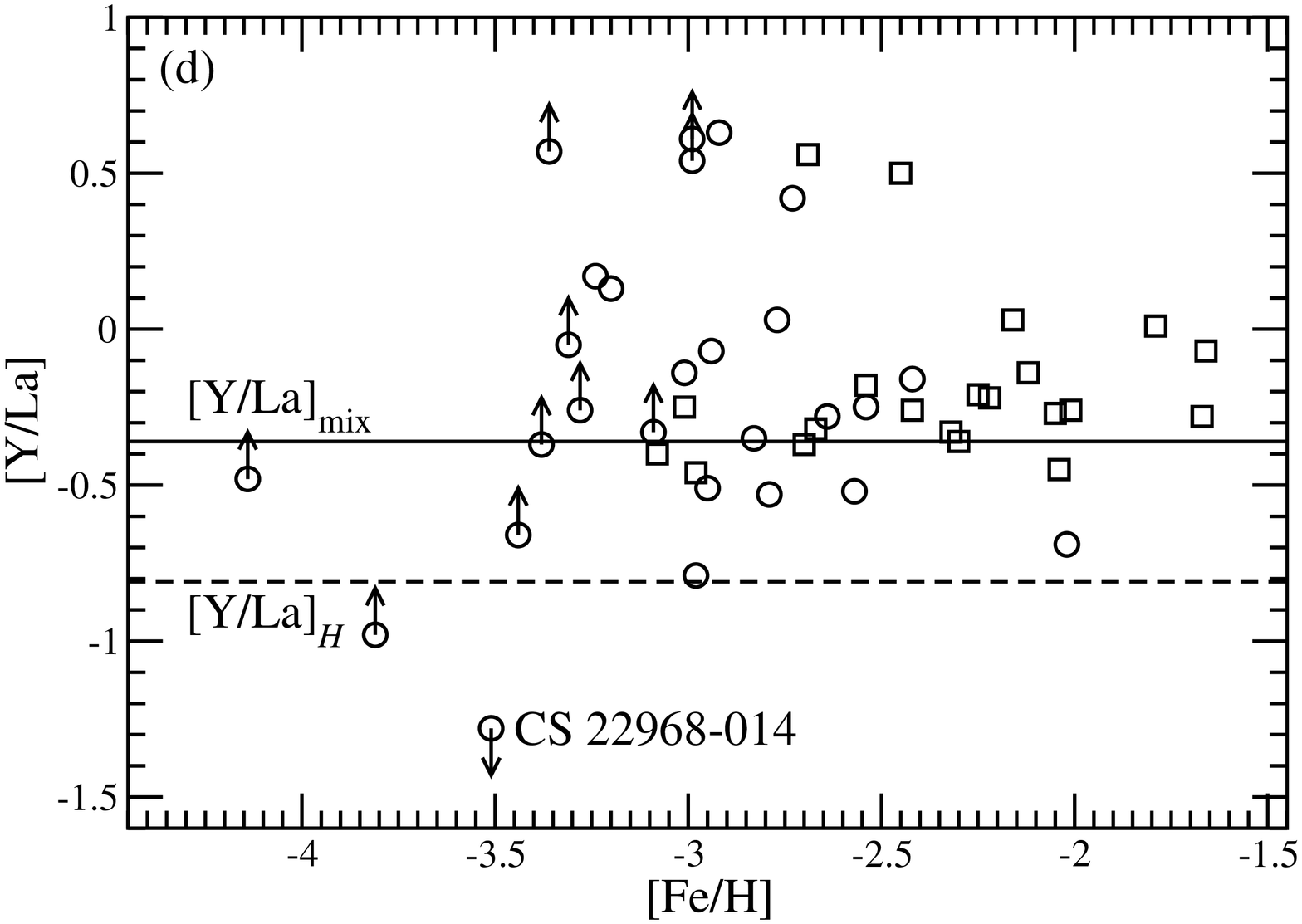}
\caption{(a) High-resolution data on $\log\epsilon({\rm Y})$ vs.
[Fe/H] (squares: \citealt{johnson}; circles: \citealt{francois07}).
Downward arrows indicate upper limits. The solid line is for an
ISM with well-mixed $H$ and $L$ contributions. Note that
many stars with [Fe/H]~$\lesssim -3$ lie below this line.
(b) Comparison of the two-component model of QW07 and
the observations in terms of $\Delta\log\epsilon({\rm Y})\equiv
\log\epsilon_{\rm cal}({\rm Y})-\log\epsilon_{\rm obs}({\rm Y})$ as a function 
of [Fe/H] for the stars shown in (a). For those stars with only
upper limits on the La abundance, only the $L$ contributions to Y are
calculated to give the lower limits on $\Delta\log\epsilon({\rm Y})$ shown
as the upward arrows. The two-component model grossly overestimates
the Y abundances at ${\rm [Fe/H]}\lesssim -3$ but describes the 
observations very well at ${\rm [Fe/H]}> -3$. (c) Evolution of [Y/Fe]
with [La/Fe] for those stars shown in (a) that have observed Y abundances.
Left-pointing arrows indicate upper limits on [La/Fe]. The curves are
calculated from the three-component model
with the $H$ and $L$ sources and a third source (HNe)
for $f_{{\rm Fe},L}=0$ (dot-dot-dashed), 0.1 (dashed), 0.5 (dot-dashed),
and 1 (solid). The filled circle labeled ``$L$''
indicates the value of [Y/Fe]$_L=-0.43$ for the $L$ source. Note that
the data mostly lie between the curves for
$f_{{\rm Fe},L}=0$ and 1. The anomalous star CS~22968--014 is
an exception. (d) Data on [Y/La] vs. [Fe/H] for the
stars shown in (c). Except for CS~22968--014,
all the other stars are consistent with the lower bound of 
${\rm [Y/La]}\geq{\rm [Y/La]}_H=-0.81$ from the three-component model.
Typical observational errors in $\log\epsilon({\rm Y})$ [see (a)], 
[Y/Fe], [La/Fe], and [Y/La] are $\sim 0.1$--0.3 dex.}
\label{fig-yla}
\end{figure}

As a final test for the requirement of a third source and the robustness 
of the extended model including this source, we repeat the analysis 
using the high-resolution data on Y and La
in metal-poor stars. Like Sr and Ba, these two elements represent
the CPR elements and the heavy $r$-elements, respectively. The
abundances of Y and La in a star are generally much lower than 
those of Sr and Ba, respectively. Consequently, there are much
fewer data on Y and La than on Sr and Ba in metal-poor stars.
On the other hand, the abundances of Y and La are less 
susceptible to uncertainties in the spectroscopic analysis if they
can be measured and therefore, may be better indicators for
the trends of chemical evolution (e.g., \citealt{simmerer}).
In Figure~\ref{fig-yla}a we show the data on $\log\epsilon({\rm Y})$
from the high-resolution observations of \citet{johnson} (squares)
and \citet{francois07} (circles) over the wide range of
$-4.1\lesssim[{\rm Fe/H}]\lesssim -1.5$. The solid line in this
figure represents
\begin{equation}
\log\epsilon({\rm Y})=[{\rm Fe/H}]+1.97,
\end{equation}
which corresponds to [Y/Fe]$_{\rm mix}=-0.24$ for an ISM
with well-mixed contributions from the $H$ and $L$ sources
only (see Appendix~\ref{sec-app1} and Table~\ref{tab-mix}). 
It can be seen
from Figure~\ref{fig-yla}a that the bulk of the data again cluster 
around the solid line, but there are again many stars with 
[Fe/H]~$\lesssim -3$ showing a great deficiency in Y.

The failure of the two-component model with the $H$ and $L$
sources as found for Sr
can also be shown by comparing the Y abundances
predicted from this model with the data on metal-poor stars. 
The yield patterns of the $H$ and $L$ sources given in 
Tables~\ref{tab-rhl} and \ref{tab-yhl} correspond to
$\log({\rm Y/La})_H=0.27$ and $\log({\rm Y/Fe})_L=-5.67$.
Using these yield ratios and the La and Fe data on the stars
shown in Figure~\ref{fig-yla}a, we calculate the Y abundances 
for these stars from
\begin{equation}
\left(\frac{\rm Y}{\rm H}\right)=\left(\frac{\rm Y}{\rm La}\right)_H
\left(\frac{\rm La}{\rm H}\right)
+\left(\frac{\rm Y}{\rm Fe}\right)_L\left(\frac{\rm Fe}{\rm H}\right)
\end{equation}
and show the differences $\Delta\log\epsilon({\rm Y})\equiv
\log\epsilon_{\rm cal}({\rm Y})-\log\epsilon_{\rm obs}({\rm Y})$
between the calculated and observed values in Figure~\ref{fig-yla}b.
As many stars lack La data, their $\log\epsilon_{\rm cal}({\rm Y})$ 
values are calculated from the $L$ contributions only. The
resulting $\Delta\log\epsilon({\rm Y})$ values represent lower
limits and are shown as symbols with upward arrows
[the value of $\log({\rm Y/Fe})_L=-5.67$ should represent the
minimum value of $\log({\rm Y/Fe})=\log\epsilon({\rm Y})-
\log\epsilon({\rm Fe})$ predicted for metal-poor stars 
based on the two-component model]. 
It can be seen from Figure~\ref{fig-yla}b that
there is again good agreement between the two-component model 
and the data for ${\rm [Fe/H]}>-3$ but the model tends to greatly 
overpredict Y abundances (by up to $\sim 1.3$ dex)
for ${\rm [Fe/H]}\lesssim -3$.

As discussed using just the Sr and Ba data, the two-component
model must be modified by including a third source producing Fe
but no CPR or heavier elements in order to account for the 
observations. The effects of such a source are shown in 
Figure~\ref{fig-yla}c for the CPR element Y and the heavy 
$r$-element La 
analogous to Figures~\ref{fig-srfe} and \ref{fig-heres}b
for Sr and Ba. The data on the evolution of 
[Y/Fe] with [La/Fe] shown in Figure~\ref{fig-yla}c are for the stars 
shown in Figure~\ref{fig-yla}a except for those with only upper 
limits on both Y and La abundances. The distribution of the
data on Y and La with respect to the curves calculated from the 
three-component model for $f_{{\rm Fe},L}=0$, 0.1, 0.5, and 1 
is similar to those discussed in \S\ref{sec-p}
for the evolution of [Sr/Fe] with [Ba/Fe] and
shown in Figures~\ref{fig-srfe} and \ref{fig-heres}b. There is
one exceptional star, CS~22968--014, which is indicated by
the downward arrow labeled as such in Figure~\ref{fig-yla}c.
This star has an anomalously high abundance of La 
corresponding to $\log({\rm La/Ba})=0.7$ \citep{francois07}, 
which greatly exceeds the yield ratio of
$\log({\rm La/Ba})_H=-0.71$ assumed 
for the $H$ source (see Table~\ref{tab-rhl}). If the measured
La/Ba ratio were correct, then CS~22968--014 must have 
sampled an extremely anomalous event producing the 
heavy $r$-elements. However, the Ba and La abundances
for this star were derived from a single line for either element
\citep{francois07}, and therefore, could be in error. More 
observations of these
two elements in this star are needed to resolve this issue.
In any case, we consider that the overall comparison between 
the theoretical model 
curves for $f_{{\rm Fe},L}=0$ to 1 and the data shown in 
Figure~\ref{fig-yla}c justifies the three-component model where
the third source is producing Fe but no CPR or heavier elements.

With no production of CPR or heavier elements assigned to the third 
source, the Y/La ratio is determined exclusively by the $H$ and $L$ 
sources. Any mixture of the contributions from these two sources 
should have ${\rm [Y/La]}$ exceeding ${\rm [Y/La]}_H=-0.81$
(see Table~\ref{tab-mix}). An ISM with well-mixed $H$ and $L$ 
contributions should have [Y/La]$_{\rm mix}=-0.36$ 
(see Appendix~\ref{sec-app1} and Table~\ref{tab-mix}). 
The data on [Y/La] for the stars 
shown in Figure~\ref{fig-yla}c are displayed in Figure~\ref{fig-yla}d
analogous to Figure~\ref{fig-srba}. It can be seen from 
Figure~\ref{fig-yla}d that except for the anomalous star 
CS~22968--014 noted above, all other data are compatible 
with the lower bound of ${\rm [Y/La]}\geq {\rm [Y/La]}_H$ and that a 
large fraction of the stars did not sample a well-mixed ISM. 

Based on the analysis of the Sr and Ba data as well as the
Y and La data, we consider that a three-component model including
a third source producing Fe but no CPR or heavier elements is 
adequately justified. Our analysis of the Zr and Ba data 
(not presented in detail here) is in full accord with the three-component
model and leads to the same quantitative conclusion
(see \S\ref{sec-3cm} and Figure~\ref{fig-csrfe}d). We will now pursue 
the consequences of this approach.

\subsection{HNe as the Third Source}
\label{sec-plowa}
Star formation in the early universe responsible 
for the enrichment of metal-poor stars is still not well understood. 
Simulations indicate that the first stars were likely to be massive, 
ranging from $\sim 10$ to $\sim 1000\,M_\odot$
[see \citet{abel02,bromm04} for reviews of earlier works and 
\citet{yoshida06,oshea07,gao07} for more recent studies]. 
It is generally thought that stars form in the typical mass range of 
$\sim 1$--$50\,M_\odot$ subsequent to the epoch of the first stars.
Below we assume this simple scenario of star formation and
focus on considerations of nucleosynthesis to identify the stellar
types for the third source.

The assumed third source produces the low-$A$ elements
including Fe but no CPR elements such as Sr or heavier elements.
As the CPR elements are here considered to
be produced in the neutrino-driven wind from nascent neutron stars, 
there are two main candidates for the third source : 
(1) pair-instability SNe (PI-SNe)
from very massive ($\sim 140$--$260\,M_\odot$) stars (VMSs), 
in which the star is completely disrupted by the explosion 
and no neutron star is produced, and (2) massive SNe with progenitors of 
$\sim 25$--$50\,M_\odot$, in which a black hole forms either
directly by the core collapse or through severe fallback onto the neutron star 
initially produced by the core collapse. There is observational evidence that 
massive SNe have two branches: HNe and faint SNe with the latter thought
to be much rarer. Compared with normal SNe,
HNe have up to $\sim 50$ times higher explosion energies and 
$\sim 7$ times higher Fe yields while faint SNe have several times lower 
explosion energies and $\gtrsim 10$ times lower Fe yields
[see \citet{iwa98} for interpretation of SN~1998bw as an HN,
\citet{tura} for the case of SN~1997D as a faint SN, and
\citet{nomoto06} and references therein for other studies of HNe
and faint SNe]. It is important to note that HNe are ongoing events in
the present universe as evidenced by the occurrences of the associated
gamma-ray bursts [see e.g., \citet{galama} for the discovery of SN~1998bw,
an HN associated with a gamma-ray burst].

In our assumed scenario of star formation, PI-SNe can only occur at
zero metallicity but HNe and faint SNe can occur at all epochs. In addition,
these three types of events have very different yield patterns of the 
low-$A$ elements. 
Compared with HNe and faint SNe, PI-SNe have extremely low production
of those low-$A$ elements with odd atomic numbers such as
Na, Al, K, Sc, V, Mn, and Co relative to their neighboring elements
with even atomic numbers (see Figure~3 in \citealt{heger02}). 
This is because unlike HNe and faint SNe
that occur after all stages of core burning, PI-SNe occur immediately 
following core C-burning and there is not sufficient time for weak 
interaction to provide the required neutron excess for significant
production of the low-$A$ elements with odd atomic numbers
(e.g., \citealt{heger02}). Further, the production of the low-$A$
elements from Na through Mg relative to those from Si
through Zn differs greatly between HNe and faint SNe. This is 
because the former elements are produced by hydrostatic burning 
during the pre-explosion evolution and the latter ones by explosive 
burning. The extremely weak
explosion of faint SNe would lead to very high yield ratios of
the hydrostatic burning products relative to the explosive burning
products.

\begin{figure}
\includegraphics[angle=270,scale=.50]{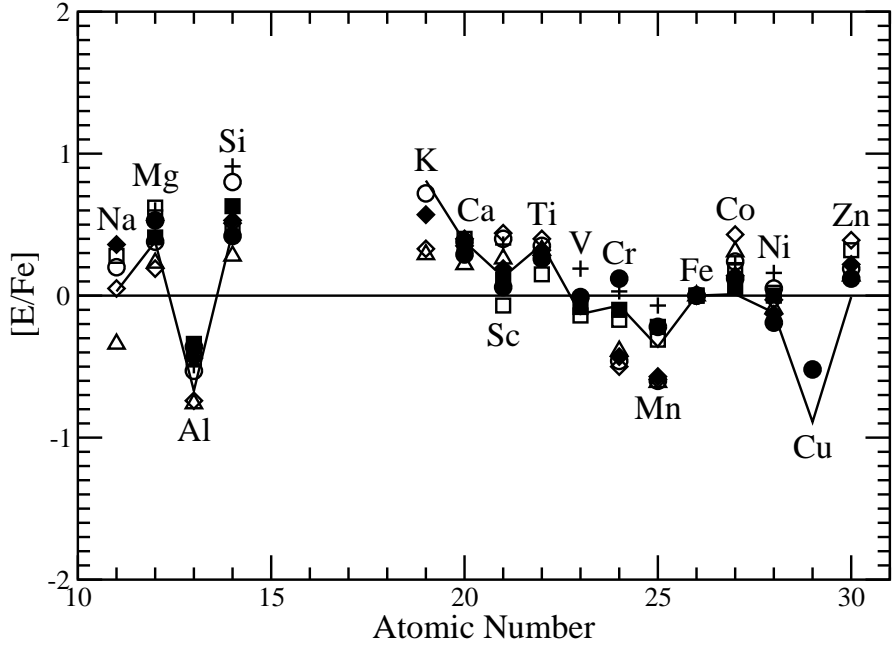}
\caption{Comparison of the abundance patterns of the low-$A$ 
elements for the third source (HNe) and the $L$ source. 
The patterns for the third source
are taken from five stars 
that lie on the curve for $f_{{\rm Fe},L}=0$ in Figure~\ref{fig-srfe}
(open square: BD~$-18^\circ 5550$, 
${\rm [Fe/H]}=-2.98$, \citealt{johnson}; open circle:
CS~30325--094, ${\rm [Fe/H]}=-3.25$, open diamond: 
CS~22885--096, ${\rm [Fe/H]}=-3.73$, open triangle:
CS~29502--042, ${\rm [Fe/H]}=-3.14$, \citealt{cayrel};
plus: BS~16085--050, ${\rm [Fe/H]}=-2.85$, \citealt{honda04}).
Those for the $L$ source are from three stars
that lie on the curve for $f_{{\rm Fe},L}=1$ in Figure~\ref{fig-srfe} 
(filled square:
BD~$+4^\circ 2621$, \citealt{johnson}; filled circle: HD~122563, 
\citealt{honda04,honda06}; filled diamond: CS~29491--053, 
\citealt{cayrel}). The solid
curve represents a star (BD~$+17^\circ 3248$, \citealt{cowan02}) 
with a relatively high value of
${\rm [Fe/H]}=-2$. Typical observational errors in [E/Fe]
are $\sim 0.1$--0.25 dex. 
All the patterns shown are essentially
indistinguishable.}
\label{fig-p}
\end{figure}

The decomposition of elemental abundances in terms of 
three components discussed in \S\ref{sec-p} and \S\ref{sec-yla}
identifies those stars in which 
the Fe is exclusively the product of the third source. Such stars lie
on the curve for $f_{{\rm Fe},L}=0$ representing the mixture of
contributions from the $H$ source and the third source  
in Figure~\ref{fig-srfe}. 
As the $H$ source produces none of the low-$A$
elements, these elements in the stars 
lying on the $f_{{\rm Fe},L}=0$ curve should be attributed to the
third source. The abundance patterns of these elements
in five such stars (open square: BD~$-18^\circ 5550$, 
${\rm [Fe/H]}=-2.98$, \citealt{johnson}; open circle:
CS~30325--094, ${\rm [Fe/H]}=-3.25$, open diamond: 
CS~22885--096, ${\rm [Fe/H]}=-3.73$, open triangle:
CS~29502--042, ${\rm [Fe/H]}=-3.14$, \citealt{cayrel};
plus: BS~16085--050, ${\rm [Fe/H]}=-2.85$, \citealt{honda04})
are shown in Figure~\ref{fig-p}.
It can be seen that all the abundance patterns of the low-$A$
elements assigned to the third source are quasi-uniform.
By quasi-uniformity, we mean that for element E, the [E/Fe] 
values for different stars are within $\sim 0.3$~dex of some
mean value. It is also clear that there are no drastic variations 
in the [E/Fe] values either between the elements with odd and 
even atomic numbers or between the hydrostatic and explosive 
burning products. We conclude that neither PI-SNe nor faint SNe 
can be the third source. This leaves HNe as the third source. 

The abundance patterns of the low-$A$ elements in those stars
that lie on the curve for $f_{{\rm Fe},L}=1$ in Figure~\ref{fig-srfe}
should represent the yield pattern of these elements for the 
hypothecated $L$ source.
The patterns for three such stars (filled square:
BD~$+4^\circ 2621$, \citealt{johnson}; filled circle: HD~122563, 
\citealt{honda04,honda06}; filled diamond: CS~29491--053, 
\citealt{cayrel}) are compared with those assigned to the third
source in Figure~\ref{fig-p}. It can be seen that the third source
(now taken to be HNe) and the $L$
source are indistinguishable in terms of their assigned
contributions to the low-$A$ elements. This is also reflected
by the fact that essentially all the stars in the region 
bounded by the curves for $f_{{\rm Fe},L}=0$ and 1 
shown in Figure~\ref{fig-srfe} have the same quasi-uniform
abundance patterns of the low-$A$ elements as
established by the observations of \citet{cayrel}
(see \S\ref{sec-pan} for discussion of the exceptional stars). 
As an example, we show in Figure~\ref{fig-p} the pattern for 
BD~$+17^\circ 3248$ (solid curve, \citealt{cowan02}) with a 
relatively high value of ${\rm [Fe/H]}=-2$.
We are thus left with a most peculiar conundrum: the yield
pattern of the low-$A$ elements attributed to the third source
is the same as that attributed to the $L$ source.
This is the same result that we \citep{qw02} found earlier in
attempting to estimate the yield patterns of the stellar sources
contributing in the regime of ${\rm [Fe/H]}\lesssim -3$
using the data of \citet{mcw} and \citet{nrb}. 
The recent more extensive and precise data of \citet{cayrel}
lead to the same conclusion. 

We have associated the third source with HNe and the $L$ 
source with normal SNe. As HNe and normal SNe are 
concurrent in our assumed scenario of star formation and 
cannot be distinguished based on their production of the
low-$A$ elements, the contributions to these elements,
especially Fe, that we previously assigned to the $L$ source 
only may well be a combination of the contributions from
both HNe and normal SNe.
In this case, the Sr/Fe ratio assigned to the $L$ source
represents a mixture of Sr contributions from normal SNe and 
Fe contributions from both HNe and normal SNe.
In what follows, we designate
normal SNe as the $L^*$ source and consider the 
$L$ source as a combination of HNe and the $L^*$ source
($L\to {\rm HNe}+L^*$). The apparent near identity in the abundance 
patterns of the low-$A$ elements attributed to HNe and the 
$L$ source may mean that the dominant
contributor to these elements is HNe.
The stellar types and the nucleosynthetic characteristics
assigned to HNe, $H$, and $L^*$ sources are summarized
in Table~\ref{tab-phl}. 

In our earlier efforts to decompose the stellar sources of
elemental abundances at low metallicities, we recognized that 
there must be a source producing Fe and other low-$A$ 
elements but none of the $r$-elements \citep{qw02}. 
We therefore proposed a source that only occurred in very 
early epochs and did not occur later. This inference, in 
conjunction with the rather sharp break in the observed 
abundances of the heavy $r$-elements at ${\rm [Fe/H]}\sim-3$, 
led us to propose that PI-SNe from VMSs might be the source. 
It was argued that VMSs were the first stars and that the very 
disruptive PI-SNe associated with them provided a baseline of 
metals to the IGM at a level of ${\rm [Fe/H]}\sim-3$. This 
apparent baseline was also found in damped Lyman $\alpha$
systems \citep{qsw}. However, in the framework of hierarchical 
structure formation, for halos that are not disrupted by explosions 
of massive stars (see \S\ref{sec-halo}), the initial rate of growth in 
metallicity is so rapid that it would be very rare to find stars with 
${\rm [Fe/H]}<-3$ \citep{qw04}. It is thus plausible that the rarity 
of ultra-metal-poor stars with ${\rm [Fe/H]}<-3$ results from the 
initial phase of rapid metal enrichment in all bound halos and is 
not due to a general ``prompt inventory'' in the IGM. In addition, 
as discussed above, none of the metal-poor stars with 
${\rm [Fe/H]}\lesssim-3$ exhibit the abundance patterns 
calculated for PI-SNe, which
are extremely deficient in the elements with odd atomic 
numbers such as Na, Al, K, Sc, V, Mn, and Co (e.g., \citealt{heger02}). 
Further, the search for ultra-metal-poor stars has shown that while 
stars with ${\rm [Fe/H]}<-3$ are rare, they do occur and show some 
evidence of elements heavier than the Fe group in their spectra 
(see \citealt{christlieb02,frebel05} for the discovery of the two most 
metal-poor stars with ${\rm [Fe/H]}<-5$). Thus, low-mass stars must 
be able to form from a medium with ${\rm [Fe/H]}\ll -3$. Based on all 
the above considerations, we now must withdraw the 
``prompt inventory'' hypothesis and must consider an IGM with
widely variable ``metal'' content and that ${\rm [Fe/H]}\sim -3$
represents a transition to the regime where halos are no
longer disrupted by the explosions of massive stars.

\section{The Three-Component Model with HNe, $H$, and $L^*$ Sources}
\label{sec-3cm}

\begin{figure}
\includegraphics[angle=270,scale=.30]{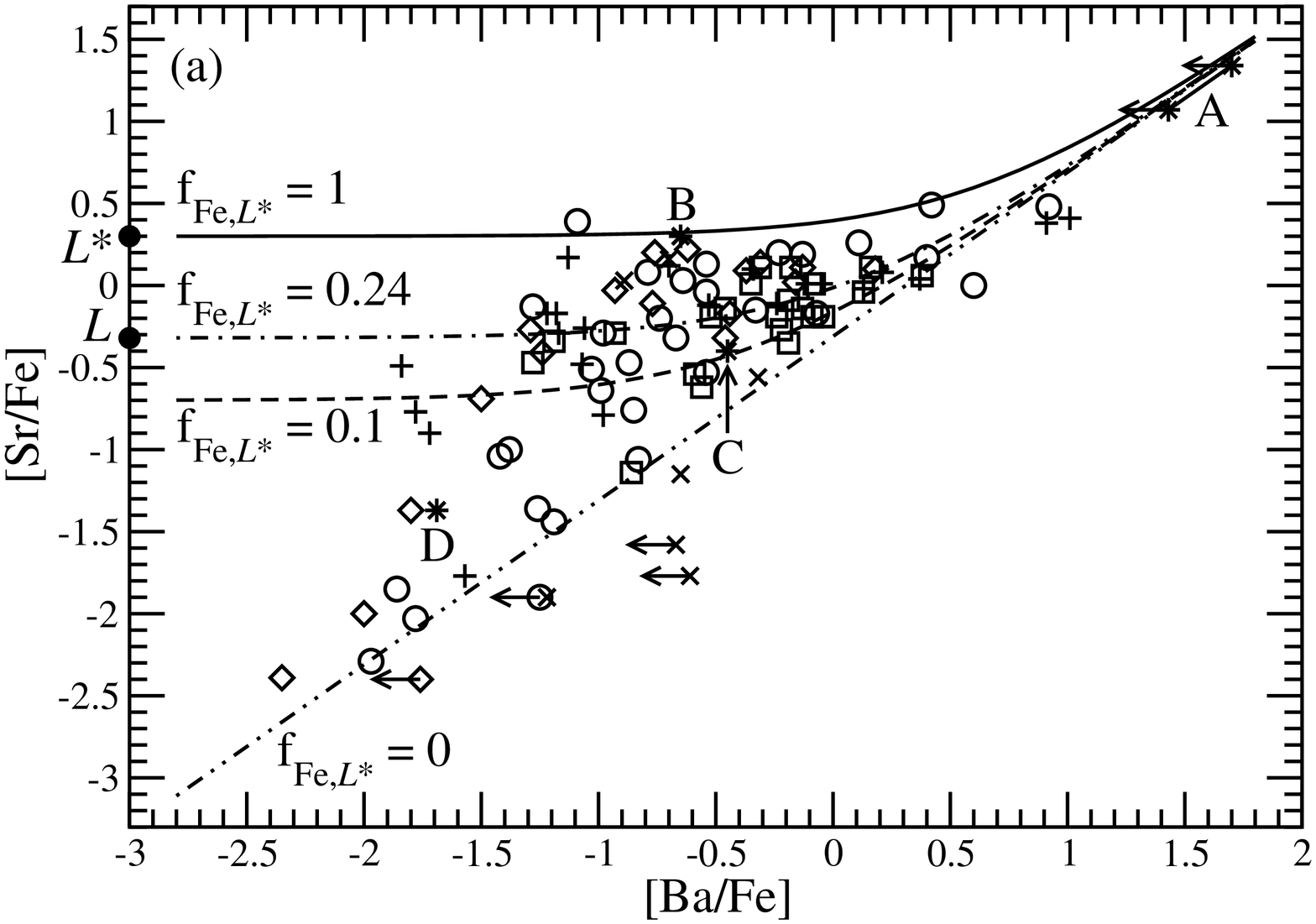}
\includegraphics[angle=270,scale=.30]{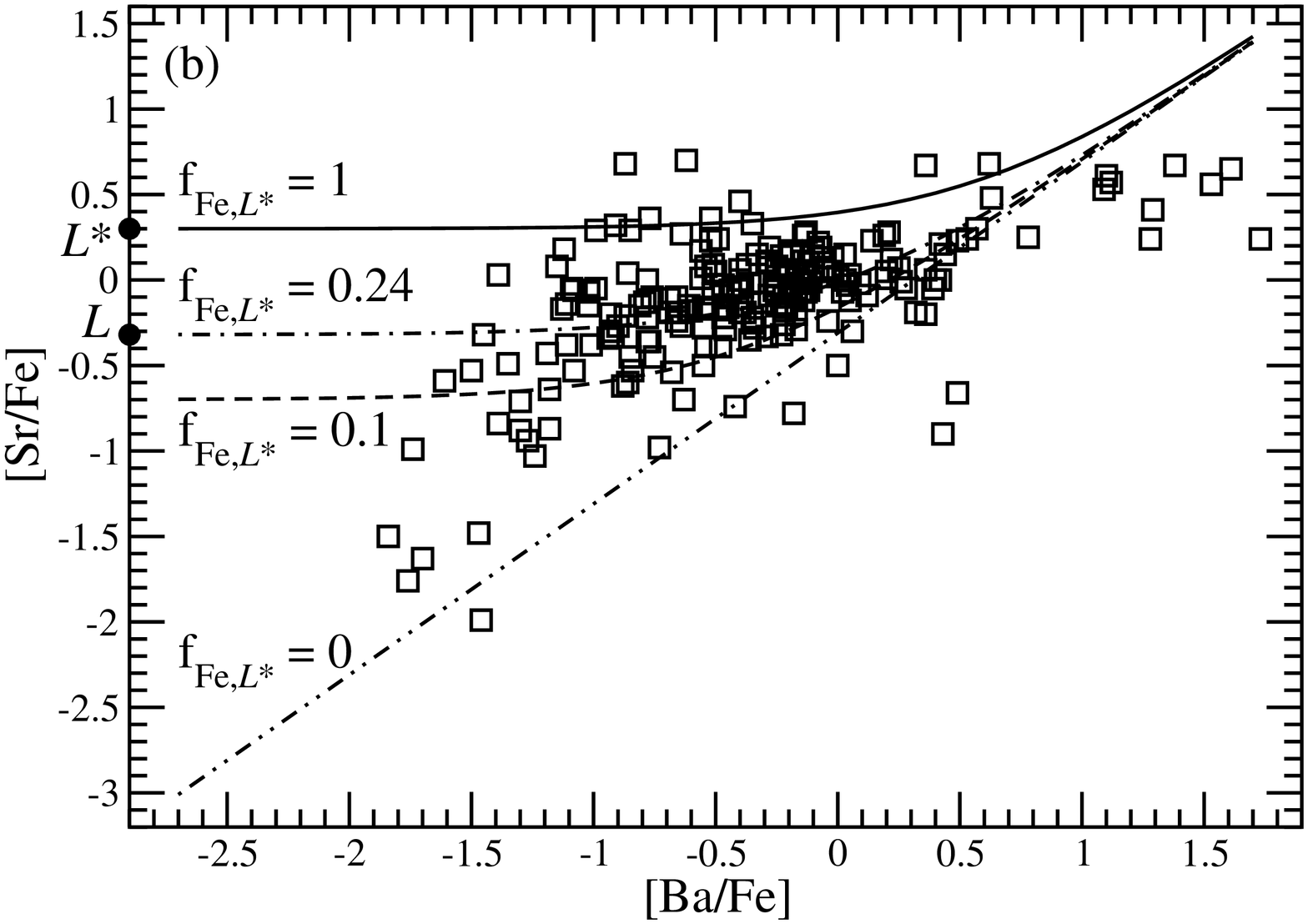}
\includegraphics[angle=270,scale=.30]{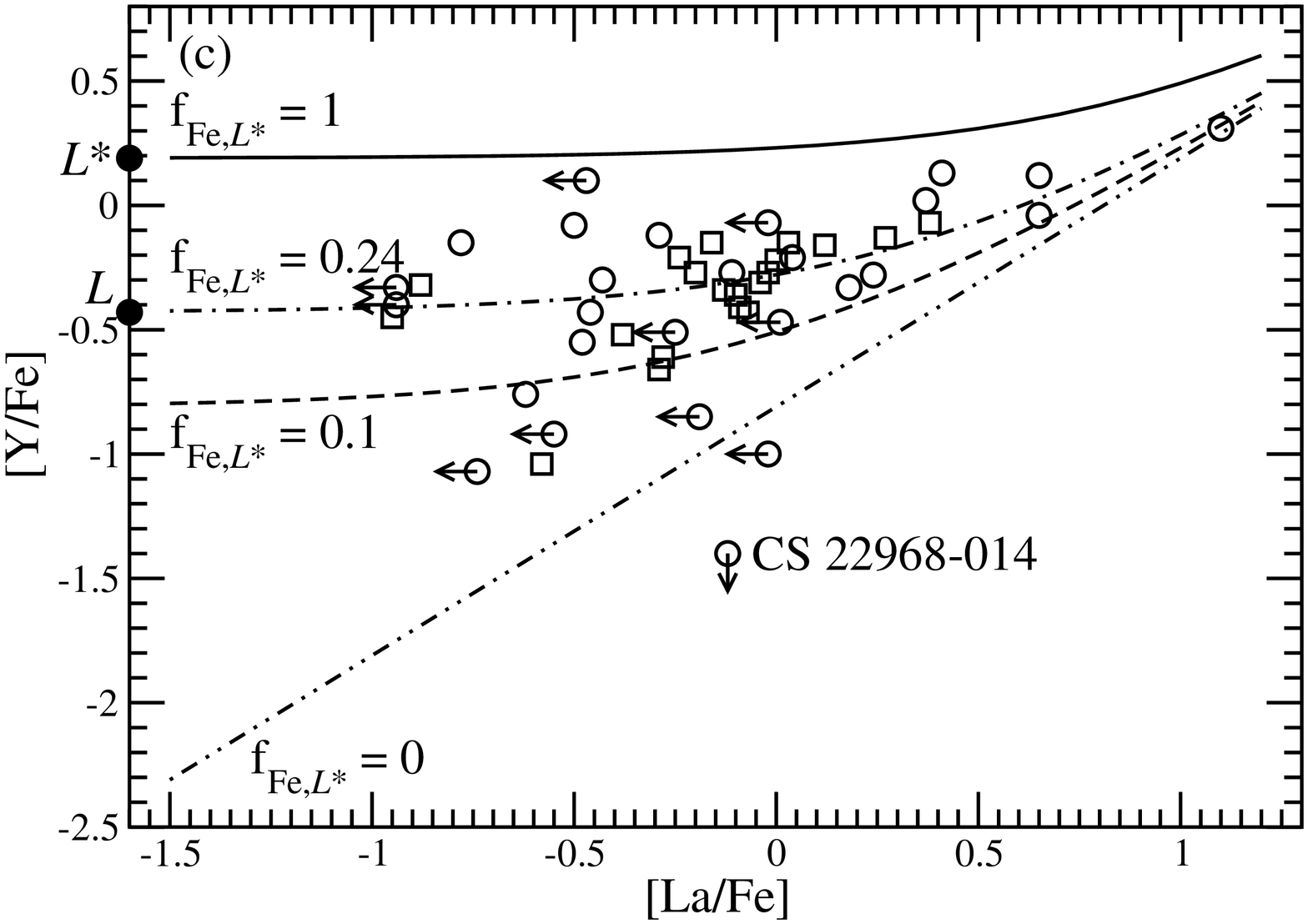}
\includegraphics[angle=270,scale=.30]{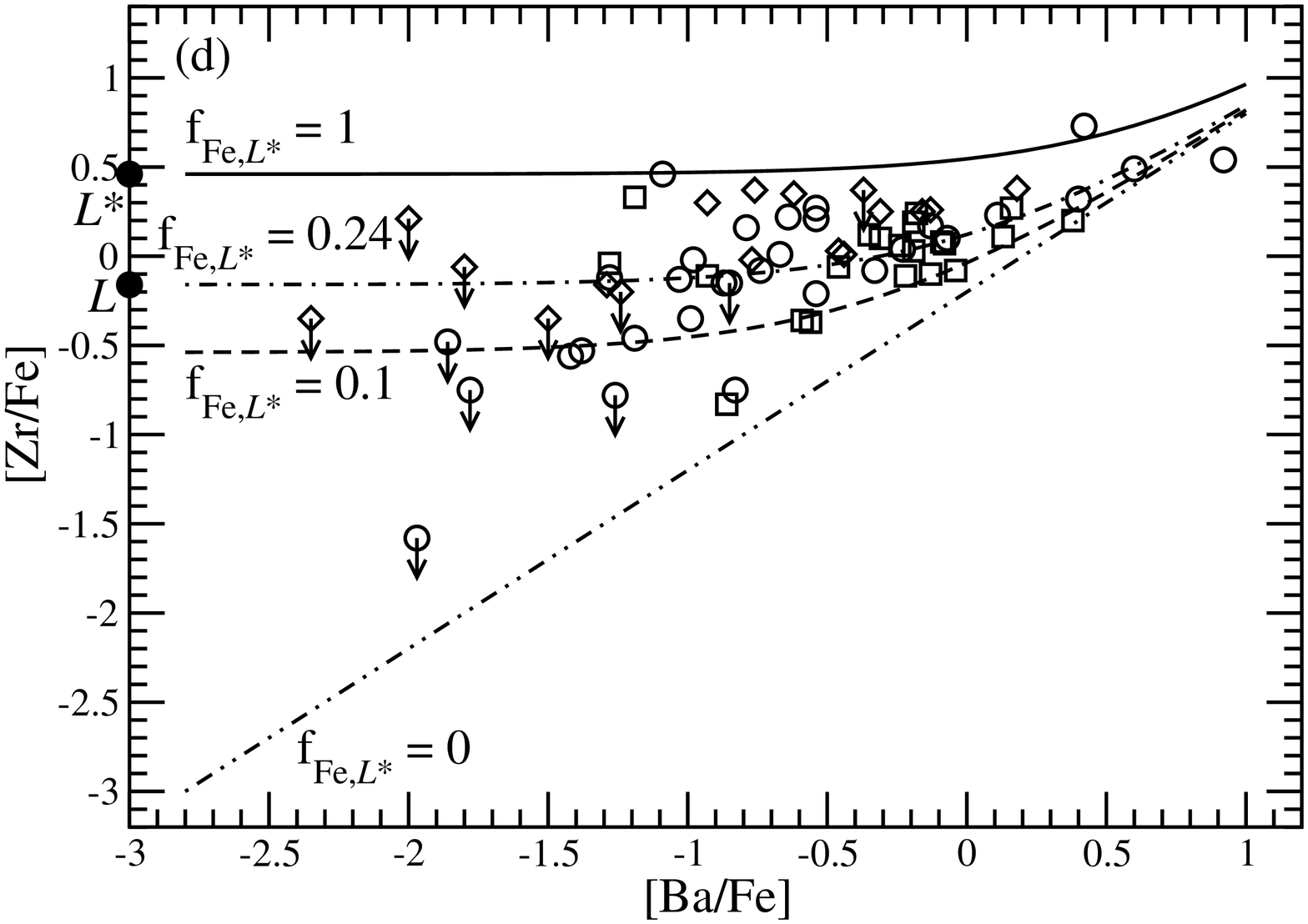}
\caption{(a) Evolution of [Sr/Fe] with [Ba/Fe] in the
three-component model with HNe, $H$, and $L^*$
sources compared with the high-resolution data
(analogous to Figure~\ref{fig-srfe}). (b) The same
relationships compared with the medium-resolution
data (analogous to Figure~\ref{fig-heres}b). 
(c) Evolution of [Y/Fe] with [La/Fe] compared with 
the high-resolution data
(analogous to Figure~\ref{fig-yla}c). (d) Evolution of 
[Zr/Fe] with [Ba/Fe] compared with the high-resolution 
data (squares: \citealt{johnson}; diamonds: \citealt{aoki05};
circles: \citealt{francois07}). Typical observational errors in 
[Zr/Fe] and [Ba/Fe] are $\sim 0.1$--0.25 dex.
The parameter $f_{{\rm Fe},L^*}$ is the fraction of Fe 
contributed by the $L^*$ source.
The filled circles labeled ``$L$'' indicate the 
(number) yield ratios of 
[Sr/Fe]$_L=-0.32$ (a) and (b), [Y/Fe]$_L=-0.43$ (c), and
[Zr/Fe]$_L=-0.16$ (d) for the $L$ source,
while those labeled ``$L^*$'' indicate the yield ratios of
[Sr/Fe]$_{L^*}=0.30$ (a) and (b), [Y/Fe]$_{L^*}=0.19$ (c), 
and [Zr/Fe]$_{L^*}=0.46$ (d) for the $L^*$ source
(see Table~\ref{tab-mix}). The increase from the $L$
to the $L^*$ yield ratio is the same for all the CPR elements.
Note that except for
the data points far to the right of and below the
curve for $f_{{\rm Fe},L^*}=0$ in (b), which may 
represent stars with large $s$-process contributions, 
and the anomalous star CS~22968--014 noted in the text,
essentially
all the data lie inside the allowed region bounded
by the curves for $f_{{\rm Fe},L^*}=0$ and 1.}
\label{fig-csrfe}
\end{figure}

With the revised interpretation of the $L$ source
as a combination of HNe and the $L^*$ source, we can
relate ${\rm [Sr/Fe]}_L=-0.32$ (see Table~\ref{tab-mix})
to the yield ratio of Sr to Fe 
for the $L^*$ source. For example, if we assume that
24\% of the Fe in the $L$ mixture is from the $L^*$ source 
(see \S\ref{sec-hne}), this corresponds to 
${\rm [Sr/Fe]}_{L^*}={\rm [Sr/Fe]}_L-\log0.24=0.30$.
Equation~(\ref{eq-srfe}) now becomes
\begin{equation}
{\rm [Sr/Fe]}=\log\left(10^{{\rm [Sr/Ba]}_H+{\rm [Ba/Fe]}}+
f_{{\rm Fe},L^*}\times 10^{{\rm [Sr/Fe]}_{L^*}}\right),
\label{eq-csrfe}
\end{equation}
where $f_{{\rm Fe},L^*}$ is the fraction of the Fe in a star
contributed by the $L^*$ source. The curves representing
the above equation for ${\rm [Sr/Fe]}_{L^*}=0.30$ and
$f_{{\rm Fe},L^*}=0$, 0.1, 0.24, and 1 are shown along 
with the data in Figures~\ref{fig-csrfe}a (high-resolution data)
and \ref{fig-csrfe}b (medium-resolution data)  
analogous to Figures~\ref{fig-srfe} and 
\ref{fig-heres}b. It can be seen from Figures~\ref{fig-csrfe}a
and \ref{fig-csrfe}b
that essentially all the data lie inside the allowed region for 
the evolution of [Sr/Fe] with [Ba/Fe] bounded by the curves for 
$f_{{\rm Fe},L^*}=0$ and 1 
(as mentioned near the end of \S\ref{sec-p},
the exceptional data points far to the right of and below
the curve for $f_{{\rm Fe},L^*}=0$ in Figure~\ref{fig-csrfe}b
most likely represent
stars that received large $s$-process contributions to Ba). 

Assuming that 24\% of the Fe in the $L$ mixture is from 
the $L^*$ source as for Figures~\ref{fig-csrfe}a 
and \ref{fig-csrfe}b, we obtain
${\rm [Y/Fe]}_{L^*}={\rm [Y/Fe]}_L-\log0.24=0.19$
(see Table~\ref{tab-mix}).
Using this yield ratio, we show the curves representing
\begin{equation}
{\rm [Y/Fe]}=\log\left(10^{{\rm [Y/La]}_H+{\rm [La/Fe]}}+
f_{{\rm Fe},L^*}\times 10^{{\rm [Y/Fe]}_{L^*}}\right)
\label{eq-cyfe}
\end{equation}
for $f_{{\rm Fe},L^*}=0$, 0.1, 0.24, and 1 along with
the data in Figure~\ref{fig-csrfe}c analogous to 
Figure~\ref{fig-yla}d. It can be seen from Figure~\ref{fig-csrfe}c
that with the exception of the anomalous star CS~22968--014
as noted in \S\ref{sec-yla}, all other data again lie inside the 
allowed region for the evolution of [Y/Fe] with [La/Fe] bounded 
by the curves for $f_{{\rm Fe},L^*}=0$ and 1. 

For completeness, we also show the high-resolution data
on the evolution of [Zr/Fe] with [Ba/Fe] (squares: \citealt{johnson};
diamonds: \citealt{aoki05}; circles: \citealt{francois07}) in 
Figure~\ref{fig-csrfe}d along with the curves representing
\begin{equation}
{\rm [Zr/Fe]}=\log\left(10^{{\rm [Zr/Ba]}_H+{\rm [Ba/Fe]}}+
f_{{\rm Fe},L^*}\times 10^{{\rm [Zr/Fe]}_{L^*}}\right)
\label{eq-czrfe}
\end{equation}
for $f_{{\rm Fe},L^*}=0$, 0.1, 0.24, and 1. In the above equation,
we take [Zr/Ba]$_H=-0.20$ and [Zr/Fe]$_{L^*}=0.46$ (see 
Table~\ref{tab-mix}). The latter yield ratio again assumes that
24\% of the Fe in the $L$ mixture is from the $L^*$
source as for Figures~\ref{fig-csrfe}a, \ref{fig-csrfe}b, and
\ref{fig-csrfe}c. It can be seen from Figure~\ref{fig-csrfe}d
that essentially all the data again lie inside the 
allowed region for the evolution of [Zr/Fe] with [Ba/Fe] bounded 
by the curves for $f_{{\rm Fe},L^*}=0$ and 1. 

Based on the comparison of the theoretical model curves and the 
data on Sr, Y, and Zr shown in Figure~\ref{fig-csrfe}, we consider 
that the three-component model with HNe,
$H$, and $L^*$ sources provides a very good description
of the elemental abundances in metal-poor stars.
For an overwhelming portion of the metal-poor stars shown in this
figure, their inventory of Fe and other low-$A$ elements received 
significant but not dominant contributions from the $L^*$
source (normal SNe) as indicated by the corresponding
low values of $f_{{\rm Fe},L^*}$. We conclude that the bulk of
the low-$A$ elements including Fe in metal-poor stars with
${\rm [Fe/H]}\lesssim -1.5$ was provided by HNe. This may
explain why wide fluctuations in the abundance patterns of
the low-$A$ elements expected from the contributions of just
a few normal SNe are not actually observed. The matter remains
as to what the detailed yield patterns of the $L^*$ source are
for the low-$A$ elements. This is not easily addressable from
the observations of metal-poor stars as the $L^*$ contributions
only constitute a small fraction of the total abundances of these
elements. It appears that we must rely on stellar model 
calculations (e.g., \citealt{ww95,cl04}) to estimate the $L^*$ yield 
patterns of the low-$A$ elements.

A straightforward application of the three-component model is
to calculate the contributions from the $H$ and $L^*$ sources
to the solar inventory of the CPR elements. Assuming that
all of the Eu in the sun was provided by the $H$ source and 
a fraction $f_{{\rm Fe},L^*}^\odot=0.08$ of the solar Fe inventory 
was provided by the $L^*$ source 
($f_{{\rm Fe},L^*}^\odot=0.24f_{{\rm Fe},L}^\odot$ with 
$f_{{\rm Fe},L}^\odot=1/3$ being the fraction contributed by
sources other than SNe Ia as usually assumed), we calculate 
the $H$ and $L^*$ contributions to a CPR element E in the sun
from
\begin{equation}
\left(\frac{\rm E}{\rm H}\right)_{\odot,HL^*}=
\left(\frac{\rm E}{\rm Eu}\right)_H
\left(\frac{\rm Eu}{\rm H}\right)_\odot+
\left(\frac{\rm E}{\rm Fe}\right)_{L^*}
\left(\frac{\rm Fe}{\rm H}\right)_\odot f_{{\rm Fe},L^*}^\odot\ ,
\label{eq-cprs}
\end{equation}
where the yield ratios (E/Eu)$_H$ and (E/Fe)$_{L^*}$ are
given in Table~\ref{tab-yhl}. We present the results in terms of
$\log\epsilon_{\odot,HL^*}({\rm E})$ in Table~\ref{tab-cprs},
where the corresponding fraction $f_{{\rm E},HL^*}^\odot$
of the solar inventory contributed by the $H$ and $L^*$ sources 
is also given. The fraction $f_{{\rm E},HL^*}^\odot$
is in approximate agreement with the fraction 
$1-f_{{\rm E},s}^\odot$ attributed to non-$s$-process
sources by \citet{arlandini99} and \citet{travaglio} for 
the elements Mo, Ru, Rh, Pd, and Ag with small to moderate 
$s$-process contributions (see Table~\ref{tab-cprs}). 
For the elements Sr, Y, Zr, and Nb 
with large $s$-process contributions, the fraction
$f_{{\rm E},HL^*}^\odot$ is a factor of $\approx 2$ larger than 
the fraction $1-f_{{\rm E},s}^\odot$ estimated by
\citet{arlandini99}. This latter result is in agreement with
what was found earlier by us \citep{qw01} and confirmed later by
\citet{travaglio}, who carried out a detailed study of
Galactic chemical evolution for the $s$-process contributions.
To calculate the fraction $1-f_{{\rm E},s}^\odot$, \citet{travaglio} 
used as input
the $s$-process yields for stars of low and intermediate masses 
with a wide range of metallicities, the formation history of these 
stars, and the mixing characteristics of their nucleosynthetic 
products with gas in the Galaxy. In contrast,
the fraction $f_{{\rm E},HL^*}^\odot$
is calculated directly from the yield templates of 
the $H$ and $L^*$ sources. These templates are taken 
from data on metal-poor stars that formed in the regime where 
there cannot be major $s$-process contributions to the ISM
and only massive stars can plausibly contribute. The only 
assumption with regard to the solar abundances used in
calculating $f_{{\rm E},HL^*}^\odot$ is the 
assignment of a fraction $f_{{\rm Fe},L^*}^\odot=0.08$ of
the solar Fe inventory to the $L^*$ source (the fraction
from this source and HNe combined is 1/3).
It appears that the results from this simple and self-consistent
approach, and hence, the assumptions used in the
three-component model, are compatible with the non-$s$-process
contributions to the solar abundances of the CPR elements. This
provides a further test of the model and does not challenge the
assignment of major Fe production by HNe as argued here.

Below we further discuss the characteristics of HNe and the $H$
and $L^*$ sources in the three-component model
and their roles in the chemical evolution of the universe.  

\subsection{Yields of HNe, $H$, and $L^*$ Sources}
\label{sec-hne}
The yields of the low-$A$ elements for HNe are not known 
although these were estimated by parameterized calculations
(e.g., \citealt{tominaga}). The Fe yields for some HNe were 
inferred from their light curves. 
Comparison of the yield patterns of the low-$A$ elements from 
various parameterized models of HNe with the abundance patterns 
observed in metal-poor stars can be found in \citet{tominaga}.
We here focus on the contributions from HNe to the Fe in the ISM.
In the regime of ${\rm [Fe/H]}\lesssim -1.5$, only HNe and normal
SNe contribute Fe. The fraction of the Fe in a well-mixed 
ISM contributed by HNe can be estimated as
\begin{equation}
\frac{\int_{25}^{50}Y_{\rm Fe}^{\rm HN}m^{-2.35}dm}
{\int_{12}^{25}Y_{\rm Fe}^{L^*}m^{-2.35}dm+
\int_{25}^{50}Y_{\rm Fe}^{\rm HN}m^{-2.35}dm}\sim 0.72,
\label{eq-fe}
\end{equation}
where we have assumed a Salpeter initial mass function (IMF) with
$m\sim 12$--25 and 25--50 (stellar mass in units of $M_\odot$)
corresponding to progenitors of normal SNe and HNe, respectively, 
and we have taken $Y_{\rm Fe}^{\rm HN}\sim 0.5\,M_\odot$ and
$Y_{\rm Fe}^{L^*}\sim 0.07\,M_\odot$ as the (mass) yields of Fe
for an HN and a normal SN, respectively (see e.g., 
Figure 1 in \citealt{tominaga} and references therein).
The fraction of the Fe contributed by normal SNe is then $\sim 0.28$.
This is close to the fraction of 0.24 assumed for the $L^*$ contribution
to the $L$ mixture and used in Figure~\ref{fig-csrfe}. As $\sim 2/3$
of the solar Fe abundance came from SNe Ia, HNe and normal SNe
contributed $\sim 24\%$ and $\sim 9\%$ of the solar Fe inventory,
respectively.

Using the Salpeter IMF and the progenitor mass ranges assumed 
in equation~(\ref{eq-fe}), we estimate
the relative rates of HNe and low-mass ($H$) and
normal ($L^*$) SNe as
\begin{equation}
R_{\rm HN}:R_H:R_{L^*}\sim\int_{25}^{50}m^{-2.35}dm:
\int_8^{11}m^{-2.35}dm:\int_{12}^{25}m^{-2.35}dm
\sim 0.36:0.96:1,
\label{eq-rphl}
\end{equation}
where we have taken the mass range for the progenitors of 
low-mass SNe to be $m\sim 8$--11.
The rate of all core-collapse SNe in the Galaxy is estimated
to be $R_{\rm SN}^G\sim 10^{-2}$~yr$^{-1}$ (e.g., \citealt{cap}).
This gives the Galactic rates of HNe and low-mass and normal
SNe as $R_{\rm HN}^G\sim 1.6\times 10^{-3}$~yr$^{-1}$,
$R_H^G\sim 4.1\times10^{-3}$~yr$^{-1}$,
and $R_{L^*}^G\sim 4.3\times10^{-3}$~yr$^{-1}$, respectively.
Assuming that HNe and normal SNe provided a total mass
$M_{\rm gas}^G$ of gas with $\sim 1/3$
of the solar Fe abundance over the period of 
$t_G\sim 10^{10}$~yr prior to the formation
of the solar system, we have
\begin{equation}
M_{\rm gas}^G\sim 
\frac{(Y_{\rm Fe}^{\rm HN}R_{\rm HN}^G+Y_{\rm Fe}^{L^*}R_{L^*}^G)t_G}
{X_{{\rm Fe},\odot}/3}\sim 3.3\times 10^{10}\,M_\odot,
\end{equation}
which is comparable to the total stellar mass in the Galactic disk
at the present time. In the above equation,
$X_{{\rm Fe},\odot}\approx 10^{-3}$ is the mass fraction of Fe in 
the sun \citep{anders}.
As low-mass SNe are the predominant source for Eu, we can 
estimate the (mass) yield of Eu for this source as
\begin{equation}
Y_{\rm Eu}^H\sim
\frac{X_{{\rm Eu},\odot}M_{\rm gas}^G}{R_H^Gt_G}
\sim 3\times 10^{-7}\,M_\odot,
\end{equation} 
where $X_{{\rm Eu},\odot}\approx 3.75\times 10^{-10}$ is
the mass fraction of Eu in the sun \citep{anders}.
Using the above Eu yield and
$\log{\rm (Sr/Eu)}_H=1.41$ (see Table~\ref{tab-yhl}),
we can estimate the (mass) yield of Sr for a single low-mass SN
(see also QW07) as
\begin{equation}
Y_{\rm Sr}^H=Y_{\rm Eu}^H
\left(\frac{\rm Sr}{\rm Eu}\right)_H
\left(\frac{A_{\rm Sr}}{A_{\rm Eu}}\right)
\sim 4.5\times 10^{-6}\,M_\odot,
\end{equation}
where  $A_{\rm Sr}\approx 88$ and $A_{\rm Eu}\approx 152$
are the atomic weights of Sr and Eu, respectively. 
The above estimate
is consistent with the amount of ejecta from the neutrino-driven
wind (e.g., \citealt{qw96}).

Using $Y_{\rm Fe}^{L^*}\sim 0.07\,M_\odot$ and 
[Sr/Fe]$_{L^*}=0.30$ [corresponding to
$\log{\rm (Sr/Fe)}_{L^*}=-4.23$, see Tables~\ref{tab-yhl}
and \ref{tab-mix}],
we can estimate the (mass) yield of Sr for a single normal SN
(see also QW07) as
\begin{equation}
Y_{\rm Sr}^{L^*}=Y_{\rm Fe}^{L^*}
\left(\frac{\rm Sr}{\rm Fe}\right)_{L^*}
\left(\frac{A_{\rm Sr}}{A_{\rm Fe}}\right)
\sim 6.5\times 10^{-6}\,M_\odot,
\end{equation}
where $A_{\rm Fe}\approx 56$ is the atomic weight of Fe. 
The above result is very close
to the Sr yield estimated for low-mass SNe and
consistent with the production of the CPR elements
in the neutrino-driven wind. 

We emphasize that we have included
the large contributions to the solar Fe inventory from HNe
in estimating the Eu and Sr yields for low-mass SNe.
This then requires that the Fe contributions from normal SNe
be reduced by a factor of $\sim 4$ from what were assumed
previously. Likewise, the Galactic rate of  
$\sim 10^{-2}$~yr$^{-1}$ usually assumed for 
normal SNe must be reduced to 
$R_{L^*}^G\sim 4.3\times 10^{-3}$~yr$^{-1}$.

\subsection{Effects of HNe, $H$, and $L^*$ Sources on
Chemical Evolution of Halos}
\label{sec-halo}
We now estimate the enrichment resulting from a single
HN or low-mass ($H$) or normal ($L^*$) SN.
In the framework of hierarchical structure formation, 
chemical enrichment depends on the mass of the
halo hosting the stellar sources and the extent to which the
gas is bound to the halo after the explosions of these sources.
In the simplest case, the gas is bound to the halo so that all
sources contribute to the evolution of metal abundances in
the halo. For these bound halos, the amount of gas to mix 
with the debris from a stellar explosion can be estimated as 
(e.g., \citealt{thornton})
\begin{equation}
M_{\rm mix}\sim 3\times 10^4E_{{\rm expl},51}^{6/7}\,M_\odot,
\label{eq-mmix}
\end{equation}
where $E_{{\rm expl},51}$ is the explosion energy in units
of $10^{51}$~erg. The explosion energy of an HN is 
inferred from the light curves to be 
$E_{\rm expl}^{\rm HN}\sim (1$--$5)\times 10^{52}$~erg
(see e.g., Figure 1 in \citealt{tominaga} and references therein),
which corresponds to 
$M_{\rm mix}^{\rm HN}\sim\mbox{(2--$9)\times 10^5\,M_\odot$}$. 
With $Y_{\rm Fe}^{\rm HN}\sim 0.5\,M_\odot$ and
$X_{{\rm Fe},\odot}\approx 10^{-3}$, this gives
\begin{equation}
{\rm [Fe/H]}_{\rm HN}\sim\log\frac{Y_{\rm Fe}^{\rm HN}}
{X_{{\rm Fe},\odot}M_{\rm mix}^{\rm HN}}
\sim\mbox{$-3.3$ to $-2.6$}
\label{eq-fehn}
\end{equation}
for enrichment of the ISM by a single HN in bound halos. 
Similarly, using $Y_{\rm Fe}^{L^*}\sim 0.07\,M_\odot$ and 
$M_{\rm mix}^{L^*}\sim 3\times 10^4\,M_\odot$ corresponding
to $E_{\rm expl}^{L^*}\sim 10^{51}$~erg, we find that a single
normal SN would result in ${\rm [Fe/H]}_{L^*}\sim -2.6$. 
As the relative rates of HNe and low-mass and
normal SNe are comparable [see equation~(\ref{eq-rphl})],
we expect that multiple types of stellar sources would be 
sampled at ${\rm [Fe/H]}>-2.6$ in bound halos. This may
explain why HNe and the $L^*$ source can be effectively
combined into the $L$ source and the two-component
model with the $H$ and $L$ sources works rather well 
at such relatively high metallicities (see Figures~\ref{fig-esr}b
and \ref{fig-yla}b). To illustrate the effects of low-mass SNe,
we consider the enrichment of Eu. Using 
$Y_{\rm Eu}^H\sim3\times 10^{-7}\,M_\odot$,
$X_{{\rm Eu},\odot}\approx 3.75\times 10^{-10}$, and
a mixing mass of $M_{\rm mix}^H\sim 3\times 10^4\,M_\odot$, 
we find that a single low-mass SN would result in
${\rm [Eu/H]}_H\sim -1.6$. This is close to the Eu abundances
observed in CS~22892--052 and CS~31082--001
with ${\rm [Fe/H]}\approx -3$ but with extremely high
enrichments of heavy $r$-elements.

The mixing mass 
$M_{\rm mix}^{\rm HN}\sim\mbox{(2--$9)\times 10^5\,M_\odot$}$
for an HN exceeds the amount of gas ($\sim 1.5\times 10^5\,M_\odot$)
in a halo with a total mass of 
$M_h\sim 10^6\,M_\odot$ (only a fraction $\approx 0.15$ in gas and 
the rest in dark matter), in which the first stars are considered to have 
formed at redshift $z\sim 20$. On the other hand, the interaction of the 
HN debris with the gas in such a halo is complicated by the gravitational
potential of the dark matter and by the heating of the gas due to the
radiation from the HN progenitor.
\citet{kitayama} studied the effects of photo-heating of the gas by 
a $200\,M_\odot$ VMS and found that with photo-heating, an explosion
with $E_{\rm expl}\gtrsim 10^{50}$~erg is sufficient to blow out all
the gas from a halo of $10^6\,M_\odot$. In contrast, without 
photo-heating, $10^{52}<E_{\rm expl}<10^{53}$~erg is required
for the same halo. The effects of the dark matter
potential are also important.
The gravitational binding energy of the gas in a halo at $z\gg 1$ 
(e.g., \citealt{barkana}) increases with the halo mass as
\begin{equation}
E_{b,{\rm gas}}\approx 2\times 10^{49}
\left(\frac{M_h}{10^6\,M_\odot}\right)^{5/3}
\left(\frac{1+z}{10}\right)\ {\rm erg}.
\end{equation}
To blow out all the gas from a halo of $3\times 10^6\,M_\odot$
requires $10^{52}<E_{\rm expl}<10^{53}$~erg
and $E_{\rm expl}>10^{53}$~erg with and without photo-heating,
respectively. The effects of photo-heating by HN progenitors of
$\sim 25$--$50\,M_\odot$ were not studied. Based on the above 
results of \citet{kitayama}, we consider it reasonable to assume
that an HN with $E_{\rm expl}\sim (1$--$5)\times 10^{52}$~erg
would blow out all the gas from a halo of 
$\sim 10^6\,M_\odot$ but a low-mass or normal or faint SN with
$E_{\rm expl}\sim 10^{51}$~erg or less would not.

\citet{greif} showed that subsequent to the blowing-out of the
gas from a halo of $\sim 10^6\,M_\odot$ by an explosion
with $E_{\rm expl}=10^{52}$~erg, collecting the debris
and the swept-up gas requires the assemblage of a much larger
halo of $\gtrsim 10^8\,M_\odot$. It is conceivable that the debris
from several or more HNe originally hosted by different halos
would be mixed and then assembled into the much larger halo.
Stars that formed subsequently from this material would have
sampled multiple HNe and have a quasi-uniform abundance 
pattern of the low-$A$ elements. The debris from a single HN
mixed with $\sim 1.5\times 10^7\,M_\odot$ of gas in a halo of
$\sim 10^8\,M_\odot$ would give ${\rm [Fe/H]}\sim -4.5$
(cf. equation~[\ref{eq-fehn}]). This is close to the lower end
of the range of [Fe/H] values for metal-poor stars.  
\citet{kitayama} showed that
even with photo-heating, to blow out all the gas from a halo of
$\sim 10^7\,M_\odot$ requires $E_{\rm expl}>10^{53}$~erg.
Consequently, after the debris from the first HNe in halos of 
$\sim 10^6\,M_\odot$ were collected into halos of 
$\gtrsim 10^8\,M_\odot$, the debris from all subsequent
stellar explosions in the larger halos would be bound to these
halos and mixed therein. As estimated above, a single HN
results in  ${\rm [Fe/H]}\sim -3.3$ to $-2.6$ and a single
normal SN results in ${\rm [Fe/H]}\sim -2.6$ for bound
halos. We therefore expect that for these halos,
multiple types of stellar sources 
would be sampled at ${\rm [Fe/H]}>-2.6$ following 
a transition regime at 
$-4.5<{\rm [Fe/H]}\lesssim -3$. Considerations of 
bound halos with gas infall and normal star formation rates 
show that a metallicity at the level of ${\rm [Fe/H]}\sim -3$ is
reached shortly after the onset of star formation in these
halos \citep{qw04}. Thus, it is
reasonable that ${\rm [Fe/H]}\sim -3$ signifies the end of
a transition regime for the behavior of abundance patterns.

The occurrences of HNe and low-mass and
normal SNe in bound halos would result in
${\rm [Sr/Fe]}=-0.10$ and ${\rm [Ba/Fe]}=-0.20$
for a well-mixed ISM (see 
Appendix~\ref{sec-app1}). As shown in Figure~\ref{fig-srfe},
many stars have ${\rm [Sr/Fe]}\sim -2.5$ to $-1$ and
${\rm [Ba/Fe]}\sim -2.5$
to $-1$. Such low values of [Sr/Fe] and [Ba/Fe] largely reflect
the composition of the IGM immediately
following the blowing-out of the gas by the first HNe in halos
of $\sim 10^6\,M_\odot$. As HNe produce the low-$A$
elements including Fe but no Sr or heavier elements, this
IGM would have no Sr or Ba if none of the debris from
the first low-mass and normal SNe escaped from halos of 
$\sim 10^6\,M_\odot$. The very low values of 
${\rm [Sr/Fe]}\sim -2.5$ to $-1$ and
${\rm [Ba/Fe]}\sim -2.5$ to $-1$ may indicate
that $\sim 1$--10\% of the debris from the first low-mass
and normal SNe escaped from their hosting halos. Alternatively,
such low [Sr/Fe] and [Ba/Fe] values could be explained by the
mixing of the IGM that fell into the halos forming at later times
with small amounts
of the debris from low-mass and normal SNe therein.

\subsection{Exceptional Stars and Faint SNe}
\label{sec-pan}
\begin{figure}
\includegraphics[angle=270,scale=.50]{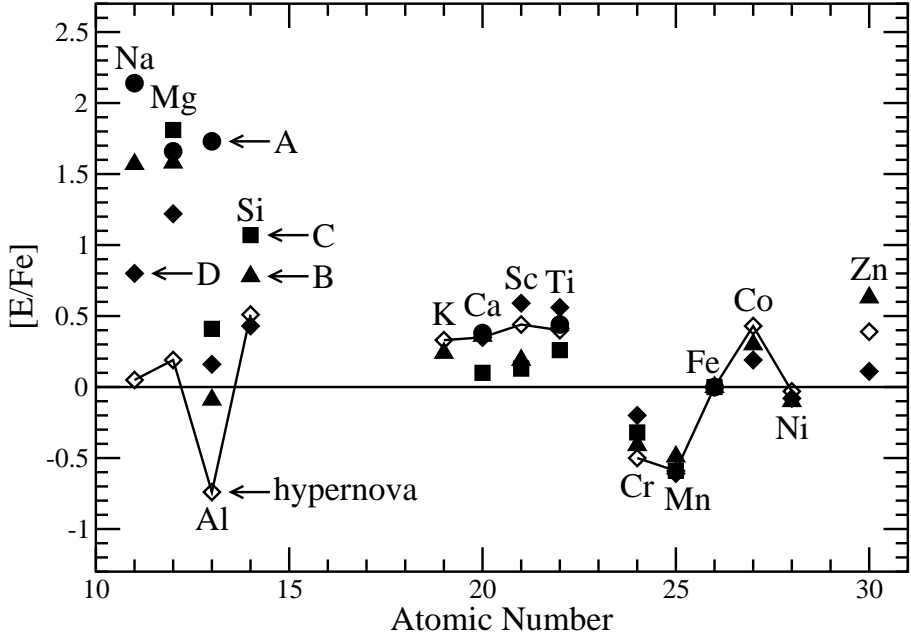}
\caption{Comparison of the abundance patterns of the low-$A$ 
elements for HNe and faint SNe. The patterns for
HNe are taken to be the same as those for the third source shown
in Figure~\ref{fig-p} and the data on CS~22885--096 
(open diamonds connected by line segments) are shown here
as a typical example.
The patterns in the anomalous stars (A, B, C, and D)
are assumed to represent faint SNe [filled circle:
\citealt{aoki06} (A, HE~1327--2326); filled triangle:
\citealt{depagne02} (B, CS~22949--037); filled square:
\citealt{aoki02} (C, CS~29498--043); filled diamond:
\citealt{aoki07} (D, BS~16934--002)]. Typical observational errors 
in [E/Fe] are $\sim 0.1$--0.25 dex. Note that the latter
patterns are characterized by extremely high abundances
of the hydrostatic burning products Na, Mg, and Al relative to
the explosive burning products from Si through Zn. Note also 
that the patterns of the explosive burning products are 
indistinguishable for HNe and faint SNe.}
\label{fig-pan}
\end{figure}

The three-component model with HNe, $H$, and $L^*$
sources describes the available data on nearly all the stars
very well. However, there are four exceptional stars that are 
identified as asterisks A, B, C, and D
in Figures~\ref{fig-esr}, \ref{fig-srfe},
\ref{fig-srba}, and \ref{fig-csrfe}a. These stars
are not anomalous in terms of Sr and Ba as shown by
the above figures. However, Figure~\ref{fig-pan} shows that
their abundance patterns of the low-$A$ elements differ greatly
from those for HNe and all the other stars (see
discussion of Figure~\ref{fig-p} in \S\ref{sec-plowa}). More
specifically, while all stars have indistinguishable patterns
of the explosive burning products from Si through Zn,
these stars have extremely high abundances of the 
hydrostatic burning products Na, Mg, and Al relative to the 
explosive burning products. Such anomalous production
patterns can be accounted for by faint SNe
(e.g., \citealt{iwamoto}), in which
fall-back coupled with a weak explosion would
hinder the ejection of the explosive burning products in 
the inner region much more than that of the hydrostatic
burning products in the outer region. Due to the weak
explosion, the debris from faint SNe would always
be bound to their hosting halos. Mixing with the debris 
from low-mass and normal SNe to some small extent
would preserve the anomalous patterns of
the low-$A$ elements and add small amounts of Sr 
and Ba to the mixture.
The stars that formed from this mixture would then 
appear as the exceptional stars discussed above.
Using $Y_{\rm Fe}\sim 4\times 10^{-3}\,M_\odot$
and $E_{\rm expl}\sim 4\times 10^{50}$~erg inferred
from the light curve of SN~1997D and
$Y_{\rm Fe}\sim 2\times 10^{-3}\,M_\odot$
and $E_{\rm expl}\sim 6\times 10^{50}$~erg
for SN~1999br (see Figure 1 in \citealt{tominaga}
and references therein),
we find that for the corresponding mixing mass
(see equations~[\ref{eq-mmix}])
a single faint SN like these two would 
result in ${\rm [Fe/H]}\sim -3.5$ (SN~1997D)
and $\sim -4$ (SN~1999br)
(cf. equation~[\ref{eq-fehn}]). These [Fe/H] values are
close to those of the exceptional stars B
(${\rm [Fe/H]}=-3.94$) and C (${\rm [Fe/H]}=-3.70$).
This is compatible with a single faint SN giving rise
to the anomalous abundance pattern of the low-$A$
elements in each exceptional star.

\section{Conclusions}
\label{sec-con}
The two-component model of QW07 with the $H$ and $L$ 
sources provided a good description of the elemental abundances
in metal-poor stars of the Galactic halo for 
$-2.7<{\rm [Fe/H]}\lesssim -1.5$. A key ingredient of that model 
is the attribution of the elements from Sr through Ag in metal-poor stars
to the charged-particle reactions in neutrino-driven winds from
nascent neutron stars but not to the $r$-process.
However, that model cannot explain the great shortfall in the
abundances of Sr, Y, and Zr relative to Fe for stars with 
${\rm [Fe/H]}\lesssim -3$.
The observations on these three CPR elements 
require that there be an early source 
producing Fe but no Sr or heavier elements. It is shown that if
such a third source is assumed, then the data can be well
explained by an extended three-component model. From considerations 
of the abundance patterns of the low-$A$ elements (from Na through Zn), 
it is concluded that this third source is most likely associated with HNe
from massive stars of $\sim 25$--$50\,M_\odot$ that do not leave behind 
neutron stars. We here consider the third source to be HNe.

It is shown that the available data on the evolution of [Sr/Fe] with [Ba/Fe],
that of [Y/Fe] with [La/Fe], and that of [Zr/Fe] with [Ba/Fe]
are well described by the extended model with HNe,
$H$ and $L$ sources, which also provides clear 
constraints on the abundance ratios that should be seen. It is further shown
that the abundance patterns of the low-$A$ elements for HNe and the $L$
sources are not distinguishable. Considering that HNe are observed to 
be ongoing events in the present universe, we are forced to conclude that
the $L$ source, which was assumed to have provided $\sim 1/3$ of the 
solar Fe inventory (the rest attributed to SNe Ia), is in fact a
combination of normal SNe (from progenitors of $\sim 12$--$25\,M_\odot$), 
which we define as the $L^*$ source, and
HNe. The net Fe contributions from HNe are found to be $\sim 3$
times larger than those from normal SNe.

Using the three-component model with HNe, $H$, and $L^*$ sources,
we obtain a very good quantitative description of essentially all the 
available data. In particular, this model provides strong constraints on 
the evolution of [Sr/Fe] with [Ba/Fe] in terms of the allowed domain for 
these abundance ratios. It gives an equally good description 
of the data when any CPR element besides Sr (e.g., Y or Zr) or any 
heavy $r$-element besides Ba (e.g., La) is used.
The model is also compatible with the non-$s$-process contributions
to the solar abundances of all the CPR elements.
The anomalous abundance patterns of 
the low-$A$ elements observed in a small number of stars
appear to fit the description of faint SNe
(e.g., \citealt{iwamoto}), which
are a rarer type of events from the same progenitor mass range as
HNe but with even weaker explosion energies and smaller Fe yields
than normal SNe (e.g.,  \citealt{nomoto06}). The anomalous abundance
patterns observed reflect the fact that faint SNe produce very little of the
Fe group elements but an abundant amount of the elements from 
hydrostatic burning in their outer shells. This gives rise to the extremely 
high abundances of Na, Mg, and Al relative to Fe observed in the 
anomalous stars. The quasi-uniform abundance patterns of the elements 
from Si through Zn in all cases (including the stars with anomalous
abundances of Na, Mg, and Al) appear to reflect some robustness in
the outcome of explosive burning that may arise from the limited range of
conditions required for such nucleosynthesis.

In this paper we used the elemental yield patterns for three 
prototypical model sources to calculate the abundances of an 
extensive set of elements (relative to hydrogen) for metal-poor
stars with ${\rm [Fe/H]}\lesssim -1.5$. As the yield patterns 
adopted for the assumed prototypical sources are taken from 
the data on two template stars, they must represent the 
results of stellar nucleosynthesis. The full version of the 
three-component model appears very successful in calculating 
the abundances of the elements ranging from Na through Pt in 
stars with ${\rm [Fe/H]}\lesssim -1.5$. In contrast to this 
phenomenological approach, there are extensive studies of 
Galactic chemical evolution (GCE) that
use the various theoretical results on the absolute yields of metals 
for different stellar types. These theoretical yields are not 
calculated from first principles, but are dependent on the 
parametrization used in the various stellar models. In those 
GCE studies, the elemental abundances for an individual star 
are not predicted. 
Instead, general trends for the elemental abundances are calculated 
assuming different sources, the rates at which they contribute, 
and a model of mixing in the ISM for different regions of the Galaxy. 
These results give a good broad description for typical elemental
abundances in the general stellar population at higher metallicities 
of ${\rm [Fe/H]}>-1.5$. This is a regime in which the observational 
data are quite convergent with only limited variability. 
However, as anticipated by \citet{gilroy} and supported by the
considerable scatter in the abundances of heavy elements observed
in stars with ${\rm [Fe/H]}\lesssim -1.5$, the chemical composition
of the ISM in the early Galaxy was extremely inhomogeneous.
For ${\rm [Fe/H]}<-2$ there are gross discrepancies between the 
observations and the smoothed model of GCE. In no case does that 
model give the elemental abundances for an individual star. 
It is our view that the simple phenomenological model used here 
permits a clearer distinction between the different stellar sources 
contributing to the ISM and the IGM at early times. This model also
gives specific testable predictions, which can be used to further
check its validity.

In conclusion, we consider that the general three-component model
with HNe, $H$, and $L^*$ sources provides a quantitative and
self-consistent description of nearly all the available data on elemental
abundances in stars with ${\rm [Fe/H]}\lesssim -1.5$. Further, HNe
may be not only explosions from the first massive stars (i.e., the 
Population III stars much sought after by many) that provided a 
very early and variable inventory to the IGM through ejection of
enriched gas from small halos, but also are important 
ongoing contributors to the chemical evolution of the universe.

\acknowledgments
We thank an anonymous reviewer for criticisms and suggestions that
greatly improve the paper.
This work was supported in part by DOE grants DE-FG02-87ER40328
(Y. Z. Q.) and DE-FG03-88ER13851 (G. J. W.), Caltech Division
Contribution 9004 (1125). G. J. W. acknowledges NASA's 
Cosmochemistry Program for research support 
provided through J. Nuth at the Goddard Space Flight Center. He also
appreciates the generosity of the Epsilon Foundation.

\appendix
\section{Abundance Ratios in a Well-Mixed ISM}
\label{sec-app1}
Using equation~(\ref{eq-eh}), we calculate the $H$ and $L$ contributions
(Sr/H)$_{\odot,HL}$ to the solar Sr abundance as
\begin{equation}
\left(\frac{\rm Sr}{\rm H}\right)_{\odot,HL}=
\left(\frac{\rm Sr}{\rm Eu}\right)_H\left(\frac{\rm Eu}{\rm H}\right)_{\odot,H}+
\left(\frac{\rm Sr}{\rm Fe}\right)_L\left(\frac{\rm Fe}{\rm H}\right)_{\odot,L}.
\label{eq-srs}
\end{equation}
As Eu is essentially a pure heavy $r$-element, we take the $H$ 
contributions to the solar Eu abundance to be 
${\rm (Eu/H)}_{\odot,H}\approx {\rm (Eu/H)}_\odot$.
Allowing for contributions from SNe Ia, we take the $L$ contributions
to the solar Fe abundance to be
${\rm (Fe/H)}_{\odot,L}\approx {\rm (Fe/H)}_\odot/3$.
Using the yield ratios (Sr/Eu)$_H$ and (Sr/Fe)$_L$ given in 
Table~\ref{tab-yhl}, we obtain
\begin{equation}
{\rm [Sr/Fe]}_{\rm mix}\equiv\log{\rm (Sr/H)}_{\odot,HL}-
\log{\rm (Fe/H)}_{\odot,L}-\log{\rm (Sr/Fe)}_\odot=-0.10,
\label{eq-srfem}
\end{equation}
which we assume to be characteristic of an ISM with well-mixed $H$ 
and $L$ contributions. Here and throughout the paper (particularly
when presenting the data from different observational studies), we have
consistently adopted the solar abundances given by \citet{ags05}.

The $H$ contributions (Ba/H)$_{\odot,H}$ to the solar Ba abundance 
can be calculated as 
\begin{equation}
\left(\frac{\rm Ba}{\rm H}\right)_{\odot,H}=
\left(\frac{\rm Ba}{\rm Eu}\right)_H\left(\frac{\rm Eu}{\rm H}\right)_{\odot,H}
\approx\left(\frac{\rm Ba}{\rm Eu}\right)_H\left(\frac{\rm Eu}{\rm H}\right)_\odot.
\end{equation}
Together with equation~(\ref{eq-srs}), this gives
\begin{equation}
{\rm [Sr/Ba]}_{\rm mix}\equiv\log{\rm (Sr/H)}_{\odot,HL}-
\log{\rm (Ba/H)}_{\odot,H}-\log{\rm (Sr/Ba)}_\odot=0.10
\end{equation}
for an ISM with well-mixed $H$ and $L$ contributions. Combining the
above equation with equation~(\ref{eq-srfem}) gives
\begin{equation}
{\rm [Ba/Fe]}_{\rm mix}={\rm [Sr/Fe]}_{\rm mix}-
{\rm [Sr/Ba]}_{\rm mix}=-0.20.
\end{equation}

Other abundance ratios such as [Y/Fe]$_{\rm mix}$, [Y/La]$_{\rm mix}$, 
[Zr/Fe]$_{\rm mix}$, and [Zr/Ba]$_{\rm mix}$ for an ISM with 
well-mixed $H$ and $L$ contributions can be calculated similarly
and are given in Table~\ref{tab-mix}.

\clearpage

\begin{deluxetable}{crrcrr}
\tablecaption{Yield ratios (E/Eu)$_H$ and (E/Fe)$_L$ for the 
heavy $r$-elements \label{tab-rhl}}
\tablewidth{0pt}
\tablehead{
\colhead{Element}&\colhead{$\log({\rm E/Eu})_H$}&
\colhead{$\log({\rm E/Fe})_L$}&\colhead{Element}&
\colhead{$\log({\rm E/Eu})_H$}&\colhead{$\log({\rm E/Fe})_L$}
}
\startdata
Ba&0.97&$-\infty$&Tm&$-0.45$&$-\infty$\\
La&0.26&$-\infty$&Yb&0.26&$-\infty$\\
Ce&0.46&$-\infty$&Lu&$-0.50$&$-\infty$\\
Pr&$-0.03$&$-\infty$&Hf&$-0.13$&$-\infty$\\
Nd&0.58&$-\infty$&Ta&$-0.88$&$-\infty$\\
Sm&0.28&$-\infty$&W&$-0.20$&$-\infty$\\
Gd&0.48&$-\infty$&Re&$-0.27$&$-\infty$\\
Tb&$-0.22$&$-\infty$&Os&0.82&$-\infty$\\
Dy&0.56&$-\infty$&Ir&0.85&$-\infty$\\
Ho&$-0.05$&$-\infty$&Pt&1.14&$-\infty$\\
Er&0.35&$-\infty$&Au&0.28&$-\infty$\\
\enddata
\tablecomments{The (number) yield ratios (E/Eu)$_H$ for the heavy
$r$-elements are taken from the corresponding solar $r$-process
abundances calculated by \citealt{arlandini99}.  The value of
$\log{\rm (Ba/Eu)}=0.96$ obtained this way is essentially the
same as the value of $\log{\rm (Ba/Eu)}=0.97$ obtained from 
the data on CS~22892--052 \citep{sneden03}. We adopt
$\log{\rm (Ba/Eu)}_H=0.97$. The yield ratios (E/Fe)$_{L^*}$ 
for the heavy $r$-elements are the same as (E/Fe)$_L$.}
\end{deluxetable}

\begin{deluxetable}{crrrr}
\tablecaption{Yield ratios (E/Eu)$_H$,  (E/Ba)$_H$, (E/Fe)$_L$,
and (E/Fe)$_{L^*}$ for the CPR elements \label{tab-yhl}}
\tablewidth{0pt}
\tablehead{
\colhead{Element}&\colhead{$\log({\rm E/Eu})_H$}&
\colhead{$\log({\rm E/Ba})_H$}&\colhead{$\log({\rm E/Fe})_L$}&
\colhead{$\log({\rm E/Fe})_{L^*}$}
}
\startdata
Fe&$-\infty$&$-\infty$&0&0\\
Eu&0&$-0.97$&$-\infty$&$-\infty$\\
Ba&0.97&0&$-\infty$&$-\infty$\\
Sr&1.41&0.44&$-4.85$&$-4.23$\\
Y&0.53&$-0.44$&$-5.67$&$-5.05$\\
Zr&1.19&0.22&$-5.02$&$-4.40$\\
Nb&0.15&$-0.82$&$-6.22$&$-5.60$\\
Mo&0.55&$-0.42$&$-5.61$&$-4.99$\\
Ru&1.03&0.06&$-5.60$&$-4.98$\\
Rh&0.40&$-0.57$&$<-5.94$&$<-5.32$\\
Pd&0.66&$-0.31$&$-6.10$&$-5.48$\\
Ag&0.07&$-0.90$&$-6.62$&$-6.00$\\
\enddata
\tablecomments{The (number) yield ratios (E/Eu)$_H$ and (E/Ba)$_H$
for the CPR elements are taken from the data on CS~22892--052
\citep{sneden03} and (E/Fe)$_L$ from the data on HD~122563
\citep{honda06}. The yield ratios
(E/Fe)$_{L^*}$ for the CPR elements are obtained from
(E/Fe)$_L$ assuming that 24\%
of the Fe in the $L$ mixture is from the $L^*$ source.}
\end{deluxetable}

\begin{deluxetable}{lrrr}
\tablecaption{Abundance and Yield Ratios Relative to Solar Values\label{tab-mix}}
\tablewidth{0pt}
\tablehead{
\colhead{Element}&\colhead{Sr}&
\colhead{Y}&\colhead{Zr}
}
\startdata
${\rm [E/Fe]}_{\rm mix}$&$-0.10$&$-0.24$&0.04\\
${\rm [E/Ba]}_{\rm mix}$&0.10&$-0.03$&0.24\\
${\rm [E/La]}_{\rm mix}$&$-0.23$&$-0.36$&$-0.09$\\
${\rm [E/Ba]}_H$&$-0.31$&$-0.48$&$-0.20$\\
${\rm [E/La]}_H$&$-0.64$&$-0.81$&$-0.53$\\
${\rm [E/Fe]}_L$&$-0.32$&$-0.43$&$-0.16$\\
${\rm [E/Fe]}_{L^*}$&0.30&0.19&0.46\\
\enddata
\tablecomments{The (number) abundance ratios with subscripts ``mix'' are
calculated for a well-mixed ISM with $H$ and $L$ contributions
only (see Appendix~\ref{sec-app1}). 
The (number) yield ratios for the $H$, $L$, and $L^*$ sources
are labeled with the corresponding subscripts. The $L^*$ yield ratios
are calculated from the $L$ yield ratios assuming that 24\%
of the Fe in the $L$ mixture is from the $L^*$ source. The solar abundances
used are taken from \citealt{ags05}.}
\end{deluxetable}

\begin{deluxetable}{cccc}
\tablecaption{Characteristics of HNe, $H$, and $L^*$ Sources
\label{tab-phl}}
\tablewidth{0pt}
\tablehead{
\colhead{Sources}&\colhead{HNe}&
\colhead{$H$}&\colhead{$L^*$}
}
\startdata
stellar types&HNe from stars&low-mass SNe from&normal SNe from\\
&of $\sim 25$--$50\,M_\odot$&stars of $\sim 8$--$11\,M_\odot$&
stars of $\sim 12$--$25\,M_\odot$\\
&&&\\
remnants&black holes&neutron stars&neutron stars\\
&&&\\
nucleosynthetic&dominant source for&source for CPR elements&
source for low-$A$\\
characteristics&low-$A$ elements&from Sr through Ag&
and CPR elements\\
&from Na through Zn&and only source for heavy&\\
&$f_{\rm Fe, HN}^\odot\sim 0.24$\tablenotemark{a}
&$r$-elements with $A>130$&
$f_{{\rm Fe},L^*}^\odot\sim 0.09$\tablenotemark{b}\\
\enddata
\tablenotetext{a}{Fraction of the solar Fe abundance contributed
by HNe.}
\tablenotetext{b}{Fraction of the solar Fe abundance contributed
by the $L^*$ source.}
\end{deluxetable}

\begin{deluxetable}{cccccc}
\tablecaption{$H$ and $L^*$ contributions to the solar inventory
of the CPR elements \label{tab-cprs}}
\tablewidth{0pt}
\tablehead{
\colhead{Element}&\colhead{$\log\epsilon_\odot({\rm E})$}&
\colhead{$\log\epsilon_{\odot,HL^*}({\rm E})$}&
\colhead{$f_{{\rm E},HL^*}^\odot$}&
\colhead{$(1-f_{{\rm E},s}^\odot)_{\rm Arlandini}$}&
\colhead{$(1-f_{{\rm E},s}^\odot)_{\rm Travaglio}$}\\
&\colhead{(1)}&\colhead{(2)}&\colhead{(3)}&\colhead{(4)}&
\colhead{(5)}
}
\startdata
Sr&2.92&2.34&0.26&0.15&0.20\\
Y&2.21&1.50&0.19&0.08&0.26\\
Zr&2.59&2.15&0.36&0.17&0.33\\
Nb&1.42&1.01&0.39&0.15&0.31\\
Mo&1.92&1.54&0.42&0.50&0.61\\
Ru&1.84&1.77&0.85&0.68&0.76\\
Rh&1.12&$>0.92$&$>0.63$&0.86&0.90\\
Pd&1.69&1.35&0.46&0.54&0.64\\
Ag&0.94&0.79&0.71&0.80&0.91\\
\enddata
\tablecomments{Column 1 gives the solar abundances of
the CPR elements from \citealt{ags05}, col. 2 gives the 
$H$ and $L^*$ contributions to the solar inventory
of these elements as calculated from the three-component
model using the yield ratios given in Table~\ref{tab-yhl},
col. 3 gives the fraction of the solar inventory
provided by the $H$ and $L^*$ sources as
calculated from cols. 1 and 2 using
$\log f_{{\rm E},HL^*}^\odot=\log\epsilon_{\odot,HL^*}({\rm E})-
\log\epsilon_\odot({\rm E})$, and cols. 4  and 5 give the fraction
contributed by processes other than the $s$-process using
the $s$-fraction $f_{{\rm E},s}^\odot$ calculated by
\citealt{arlandini99} and \citealt{travaglio}, respectively. 
As the exact value of (Rh/Fe)$_{L^*}$ 
is unknown (see Table~\ref{tab-yhl}), we calculate only the $H$ 
contribution to Rh and give the corresponding lower limits 
in cols. 2 and 3. Note that the fraction $f_{{\rm E},HL^*}^\odot$
contributed by the $H$ and $L^*$ sources is in approximate 
agreement with the fraction $1-f_{{\rm E},s}^\odot$
attributed to non-$s$-process
sources by \citealt{arlandini99} and \citealt{travaglio} for 
the elements Mo, Ru, Rh, Pd, and Ag with small to moderate 
$s$-process contributions. For the elements Sr, Y, Zr, and Nb 
with large $s$-process contributions, the fraction
$f_{{\rm E},HL^*}^\odot$ is a factor of $\approx 2$ larger than 
the fraction $1-f_{{\rm E},s}^\odot$ estimated by
\citealt{arlandini99} but in good agreement with that estimated 
by \citealt{travaglio}, who carried out a detailed study of
Galactic chemical evolution for the $s$-process contributions
to these elements. The problem in estimating the non-$s$-process
contributions to these elements has been discussed by 
\citealt{qw01} and \citealt{travaglio}. It appears that the increase 
in the non-$s$-process contributions found again here is justified 
in terms of both abundance data on metal-poor stars and 
uncertainties in modeling the $s$-process contributions.}
\end{deluxetable}

\clearpage

\end{document}